\documentclass[conference,letterpaper]{IEEEtran}

\usepackage[utf8]{inputenc} 
\usepackage[T1]{fontenc}
\usepackage{url}
\usepackage{ifthen}
\usepackage[cmex10]{amsmath} 

\usepackage{tikz}
\usetikzlibrary{arrows.meta,decorations.pathmorphing,positioning,backgrounds,fit,calc,shapes.geometric}
\usepackage{pgfplots}
\usetikzlibrary{shadows} 
\usetikzlibrary{patterns}
\pgfplotsset{compat=newest}

\usepackage{amsmath, amssymb,scalefnt,setspace}
\usepackage{mathrsfs}
\usepackage{amsfonts}

\usepackage[IEEEtran,notheorems,hyperref]{research19}

\usepackage{xargs}
\usepackage{url}
\usepackage[caption=false,font=footnotesize]{subfig}
\usepackage{stfloats}

 \newtheorem{theorem}{Theorem}
   \newtheorem{lemma}{Lemma}
   \newtheorem{corollary}{Corollary}
    \newtheorem{definition}{Definition}
   \newtheorem{remarks}{Remarks}
   
    \newtheorem{example}{Example}

\newcommandx{\strong}[3][2=\epsilon,3=n]{\mathcal{A}^{* (#3)}_{#2} (#1)}
\newcommandx{\weak}[3][2=\epsilon,3=n]{\mathcal{A}^{ (#3)}_{#2} (#1)}
\newcommandx{\typen}[2][2=n]{\top ^{ (#2)}_{#1}}
\newcommandx{\contypen}[3][3=n]{\top ^{ (#3)}_{#1}(#2)}

\newcommandx{\alltypen}[2][2=n]{\set P_{#2}({#1})}
\newcommandx{\allcontypen}[2][2=n]{\set P_{#2}({#1})}
\newcommandx{\typeseqn}[2][2=n]{\top ^{ (#2)}({#1})}
\newcommandx{\allprob}[1]{\set P({#1})}

\let\vec\relax
\newcommand{\vec}{\mathbf}
\let\set\relax
\newcommand{\set}{\mathcal}

\newcommand{\alphabetX}{\set X}
\newcommand{\alphabetY}{\set Y}

\newcommand{\alphabetN}{\set N}
\newcommand{\setRi}{{\set R}_i}
\newcommand{\setRo}{{\set R}_o}
\newcommand{\alphabetT}{\set T}
\newcommand{\tU}{\tilde{U}}
\newcommand{\dtU}{\tilde{\tU}}
\newcommand{\tX}{\tilde{X}}
\newcommand{\tY}{\tilde{Y}}
\newcommand{\Crelay}{{\set C}_{\mbox{r}}}

\newcommand{\bfyhat}{{\mathbf{\hat  y}}}
\newcommandx{\repbfyhat}[1][1=P]{\bfyhat_{#1}}
\newcommandx{\optrepbfyhat}[1][1=P]{\bfyhat^*_{#1}}

\newcommand{\bml}[1]{\mbox{\boldmath $ #1 $}}

\newcommand{\olsi}[1]{\,\overline{\!{#1}}}
  
 \newcommand*{\QEDB}{\null\nobreak\hfill\ensuremath{\square}}%

\newcommand{\ben}{\begin{enumerate}}
\newcommand{\een}{\end{enumerate}}
\newcommand{\bi}{\begin{itemize}}
\newcommand{\ei}{\end{itemize}}

\newcommand{\limn}{\lim_{n\rightarrow\infty}}

\newcommand{\one}{\frac{1}{n}}

\newcommand{\dist}{\mathsf D}

\newcommand{\etaln}{\eta_{\mbox{\scriptsize{ln}}}}
\newcommand{\etalnt}{\tilde{\eta}_{\mbox{\scriptsize{ln}}}}
\newcommand{\etamc}{\eta_{\mbox{\scriptsize{mc}}}}
\newcommand{\etakl}{\eta_{\mbox{\scriptsize{kl}}}}

\newcommand{\etamct}{\tilde{\eta}_{\mbox{\scriptsize{mc}}}}
\newcommand{\etat}{\tilde{\eta}}
\newcommand{\etate}{\tilde{\eta}^{\mbox{\scriptsize{e}}}}
\newcommand{\IchannX}{W_{{\cal X}}^{\mbox{\scriptsize{I}}}}
\newcommand{\etalnset}{\mbox{$\eta$-LN}}
\newcommand{\etamcset}{\mbox{$\eta$-MC}}
\newcommand{\etalnsetp}[1]{\mbox{$ #1 $-LN}}
\newcommand{\etamcsetp}[1]{\mbox{$ #1 $-MC}}

\newcommand{\geqln}[1]{\succcurlyeq_{ #1 \mbox{\scriptsize{ln}}}}

\newcommand{\notgeqln}[1]{\!\not{\!\!\succcurlyeq}_{ #1 \mbox{\scriptsize{ln}}}}
\newcommand{\geqmc}[1]{\succcurlyeq_{ #1 \mbox{\scriptsize{mc}}}}

\newcommand{\dhat}[1]{\hat{\hat{ #1}}}

\newcommandx{\admchannel}[1][1=\dist]{\set W^{\leq #1} }
\newcommandx{\admchanneln}[1][1=n]{\set W^{\leq \dist}_{#1} }
\newcommandx{\discontypen}[2][2=\dist]{\set W^{\leq #2}_n ({#1})}

\newcommandx{\admchannelpermn}[1][1=n]{\bar {\set W}^{\leq \dist}_{#1} }

\begin{document}
\title{Degradedness Under Cooperation} 

\author{%
  \IEEEauthorblockN{Yossef Steinberg}
  \IEEEauthorblockA{Department of ECE \\
                    Technion - IIT\\
                    Haifa 3200003, ISRAEL\\
                    Email: ysteinbe@technion.ac.il}
}

\maketitle

\begin{abstract}
We study cooperation problems in broadcast and relay networks, 
where the receivers do not satisfy the classical physical degradedness assumptions.
New notions of degradedness, \emph{strongly less noisy} and \emph{strongly more capable}
are introduced. We show that under these conditions, 
decode and forward (D\&F) is optimal for classes of cooperative systems
with limited conference rates, thus yielding new capacity results for 
these systems.
In  particular, we derive bounds on the capacity region of a class 
of broadcast channels with cooperation, that are tight on part of the capacity region. 
It is shown that the cut-set bound is tight for classes of primitive relay
and diamond channels, beyond the physically or stochastically degraded models.
\end{abstract}

\begin{IEEEkeywords}
Conference links, cooperative broadcast channels, degraded channels, diamond channels,
primitive relay, strong data processing inequalities.
\end{IEEEkeywords}

\section{Introduction}\label{sec:intro}
The concept of degraded users plays a key role in 
the analysis of communication networks with many receivers.
When the receivers can be ordered according to their channel quality,
the stronger receiver can decode the messages that the weaker receiver can, 
independently of the coding scheme used.
While well understood in the context of broadcast channels (BCs),
the notion of degradedness becomes quite subtle 
once cooperation links are introduced between receivers.
Under cooperation, the quality of measurements available at a given site 
depends on the channel, and on the signals sent from other users,
thus also on their coding schemes. 
This raises the question of how to define and quantify
degradedness in cooperative systems, 
beyond the stringent physical degradedness condition.

To set ideas, let us briefly overview degraded BCs.
The capacity of the BC is still unknown in general. However, restricted families of models 
have been solved. These include mainly physically and stochastically degraded channels.
A key observation in the study of the general
BC is that its capacity
region depends only on the channel conditional marginals. Thus, for the purpose of capacity analysis,
a stochastically degraded (SD) channel is equivalent to its physically degraded (PD) version,
and they are commonly referred to as degraded BCs (DBCs). 
The less noisy (LN) and more capable (MC)~\cite{KornerMarton:77p1},~\cite{ElGamal:79p}
and the \emph{essentially} LN and MC~\cite{Nair:10p}
are weaker forms of degradedness, that do not possess a physically degraded version.
A property common to all the degraded models mentioned above is the optimality
of superposition coding:
if there is an
order between the users, layered coding does not inflict penalty, since anyway
the stronger user can decode the messages of the weaker one. 
Superposition coding achieves capacity also for the two users  BC 
with degraded message sets~\cite{KornerMarton:77p2} - termed as asymmetric BC (ABC) 
in~\cite{CsiszarKorner:82b} - 
although the channel distribution does not imply order between the users. 
Here the degradation is present due to the problem definition:
user 1 is required to decode both private and common messages,
while user 2 decodes only the common message. 
Formal definitions of these families of channels,
and their respective capacity regions, can be found in~\cite{ElGamalKim:11b}.

The situation changes fundamentally once cooperation links between users are introduced.
A cooperation link enables one user to send a compressed version of its output to  the other user,
thus making their joint distribution relevant.
Consequently, under cooperation the capacity  region
of a stochastically degraded BC does not coincide with that of its physically degraded version,
and the models are studied separately.
Thus far, only the physically degraded BC and the ABC are solved 
under cooperation~\cite{LiangVeeravalli:07p,DaboraServetto:06p,Steinberg:15c,ItzhakSteinberg:21p}. 
In these works it is shown that the optimal strategy for this class of channels 
consists of superposition coding at the  transmitter,
and decode and forward (D\&F) at the stronger user. 
It should be noted, however, that in practical communication networks, 
the physical degradedness assumption is often non-realistic.

This work focuses on 
stochastically degraded, less noisy and more capable cooperative systems,
with cooperation links from the stronger
user to the weaker, where D\&F can be useful.
One difficulty that arises in the study of these problems, is that  
the order of the users 
is not necessarily kept under cooperation. In fact, depending on the cooperation rate, the order can be flipped. 
This is in contrast to the physically degraded channel and the ABC, 
where cooperation cannot alter the order of users: the former due to the Markov structure, and the latter due to
the inherent degradation imposed by the problem definition. 
Engineering insights suggest that if the difference between the channels
is large enough (in some sense) compared to the cooperation rate, the order of users
should not be altered, independently of the coding and cooperation strategies.
But thus far, all the definitions of degradedness relations are binary: 
a channel is either (physically, stochastically, less noisy, more capable) degraded, or not.
Therefore, a first step would be to \emph{quantify} degradedness relations, in a way
that they can be compared to cooperation rates. In the current work we put these ideas
on a firm ground, and provide quantitative degradedness conditions
under which cooperative systems can be solved.
The new definitions are extensions of 
K\"{o}rner and Marton's LN and MC conditions~\cite{KornerMarton:77p1}, 
and strong data processing inequalities 
(SDPIs)~\cite{AhlswedeGacs:76p,Raginsky:16p,PolyanskiyWu:17p}.

The primitive relay channel (PRC) is a special case of the BC with cooperation.
It should be noted that thus far only physically degraded relay channels were defined in the literature. 
Stochastically degraded PRC were not defined and studied as separate models.
Due to the similarity between the models, our results are relevant
also to a class of stochastically degraded primitive relay channels,
and to classes of primitive relay networks, like the diamond channel (DC).
The new classes of cooperative systems inherit a few desirable properties
from the BC without cooperation:
\begin{itemize}
\item The class definitions and the results depend on the BC,
         PRC and DC only via their
         conditional marginals         
\item The order of users is kept under cooperation
\end{itemize}
In stochastically degraded channels where the cooperation direction is from the weaker user to the stronger,
D\&F cannot increase capacity, since even without the help,
the stronger user can decode the messages forwarded by the weaker user. 
Therefore in this case some form of compress and forward (C\&F) should be employed.
This class of problems is studied elsewhere.

In~\cite{FarsaniYu:25p}, inner and outer bounds are derived for general BC with bidirectional cooperation.
The outer bound is established using multiple applications of Csisz\'{a}r's sum lemma. 
For the inner bound, the conference rates are split between D\&F and C\&F strategies,
where the split parameter is subject to optimization. The bounds coincide - thus characterizing the capacity region -
for semi-deterministic (SeD) ABC, SeD BC with one-sided cooperation from the weaker to the
stronger user and Gaussian channels with perfectly correlated noise components at the receivers.
In addition, for Gaussian channels with one sided cooperation from the stronger user  to
the weaker, it is shown that the gap between the D\&F region and capacity 
is at most $\frac{1}{2}\log(2/(1-|\lambda|))$ bits,
where $\lambda$ is the correlation coefficient between the noise at the two channels. 
While interesting theoretically, SeD channels, and perfectly correlated noise at the receivers, 
are not common in practical communication systems, 
thus the applicability of the respective capacity results
 is somewhat restricted. The bounds developed are relevant to the
 cooperative BC and PRC studied here, however, the two works differ in their goals. Here we aim to
 clarify the notion of degradation in cooperative systems, and
 identify the cases where degradation relations are preserved and simple D\&F
 schemes achieve capacity. 
 In contrast, degradation relations are not used in~\cite{FarsaniYu:25p}.

The structure of this work is as follows. In Section~\ref{sec:pre}
we define new forms of degradedness - \emph{strongly less noisy}
and \emph{strongly more capable}, which
are extensions of the classical LN and MC conditions, 
in the spirit of SDPIs.
In Section~\ref{sec:main_BC} the main result for the broadcast channel with cooperation is presented
and discussed. Our results for the PRC and relay networks (diamond channels) are
presented in sections~\ref{sec:PRC} and~\ref{sec:diamond}, respectively.
Nonlinear degradedness conditions are presented briefly in Section~\ref{sec:nonlinear}.
 Finally, the proofs are given in Section~\ref{sec:proofs}. 
 The examples are derived in the Appendix.

\section{Strong degradedness relations}
\label{sec:pre}
Let $P_{Y_1|X}$, $P_{Y_2|X}$ be a pair of channels with output alphabets  
$\alphabetY_1$, $\alphabetY_2$, respectively, and common input alphabet $\alphabetX$.
Assume throughout that $C_2\eqdef\max_{P_X} I(X;Y_2)>0$.
For a given distribution $P$ on $\alphabetX$, we denote by $P_{Y|X}\circ P$
the distribution on $\alphabetY$ induced by $P$ at the input.
Let $\eta$ be a real number in $(0,1)$. We define the following extensions of 
the less noisy and more capable properties.
\begin{definition}
Let $\left\{ P_{Y_1|X},P_{Y_2|X}\right\}$ be a pair of channels with the same input alphabet and $C_2>0$. Then
\label{def:LNMC}
\begin{enumerate}
\item
$P_{Y_1|X}$ is said to be  strongly less noisy than $P_{Y_2|X}$, denoted $P_{Y_1|X}\geqln{\eta} P_{Y_2|X}$,
if 
\begin{equation}
\eta I(U;Y_1) \geq I(U;Y_2) \label{eq:sLN1}
\end{equation}
for all $P_{UX}$ such that $U\markov X\markov (Y_i)$, $i=1,2$, are Markov chains. 
The minimal~$\eta$ for which~(\ref{eq:sLN1}) holds is denoted by $\eta_{\mbox{\scriptsize{ln}}}(P_{Y_1|X},P_{Y_2|X})$.
\item
$P_{Y_1|X}$ is said to be  strongly more capable than $P_{Y_2|X}$, denoted $P_{Y_1|X}\geqmc{\eta} P_{Y_2|X}$,
if 
\begin{equation}
\eta I(X;Y_1) \geq I(X;Y_2) \label{eq:sMC1}
\end{equation}
for all $P_X$.
The minimal~$\eta$ for which~(\ref{eq:sMC1}) holds is denoted by $\eta_{\mbox{\scriptsize{mc}}}(P_{Y_1|X},P_{Y_2|X})$.
\end{enumerate}
\end{definition}
To stress the dependence on the parameter $\eta$, we will sometimes use the nomenclature $\eta$-LN (resp. $\eta$-MC)
to denote the classes of channels that are strongly less noisy (resp. strongly more capable) with parameter $\eta$.
For $\eta=1$, (\ref{eq:sLN1}) and~(\ref{eq:sMC1}) reduce to the classical definitions of less noisy and more capable relations.
When clear from the context, we omit the dependence of the minimal $\eta$'s on the conditional marginals
$P_{Y_i|X},$ $i=1,2$, and use the notation $\etaln$ and $\etamc$. By their definitions, $\etaln$ and $\etamc$
can be expressed as
\begin{IEEEeqnarray}{rCl}
\etaln &=& \sup_{P_{UX}}\frac{I(U;Y_2)}{I(U;Y_1)} \label{eq:sup_LN}\\
\etamc &=& \sup_{P_X}\frac{I(X;Y_2)}{I(X;Y_1)}\label{eq:sup_MC}
\end{IEEEeqnarray}
where the supremum in~(\ref{eq:sup_LN}) (resp.~(\ref{eq:sup_MC})) is taken over all $P_{UX}$
such that $I(U;Y_1)>0$ (resp. $I(X;Y_1)>0$). 
Thus $\etamc\leq\etaln$, and the sets are nested, i.e.,
$\eta$-LN $\subseteq\eta'$-LN $\forall\eta'\geq\eta$, and similarly
for $\eta$-MC channels.
The next statement suggests a simple criterion
to verify whether a pair $\left\{ P_{Y_1|X}, P_{Y_2|X}\right\}$ satisfies the SLN condition,
in the spirit of a parallel result for the classical LN condition~\cite[Theorem 2]{vanDijk:97p}.
\begin{lemma}
\label{lemma:LNequiv}
A pair  $\left\{ P_{Y_1|X}, P_{Y_2|X}\right\}$ is $\eta$-LN if and only if
\begin{equation}
J(\eta,P_X) \eqdef \eta I(X;Y_1) - I(X;Y_2) \label{eq:J}
\end{equation}
is concave in $P_X$.
\end{lemma}
\noindent
{\bf Proof:} We can write:
\begin{IEEEeqnarray}{rCl}
  \IEEEeqnarraymulticol{3}{l}{
\eta I(U;Y_1) - I(U;Y_2)
  = \eta I(UX;Y_1) - I(UX;Y_2) }\nonumber\\* \quad
  & & - \left[\eta I(X;Y_1|U) - I(X;Y_2|U)\right]\nonumber\\
                                       &=&  \eta I(X;Y_1) - I(X;Y_2)\nonumber\\
                                        & & - \left[\eta I(X;Y_1|U) - I(X;Y_2|U)\right],\label{eq:LNequiv}
\end{IEEEeqnarray}
and the assertion follows since $I(X;Y_i|U)$ is a convex combination of $I(X_u;Y_i)$ according to the distribution of $U$,
for $i=1,2$.
\QEDB

The strong LN condition is related to SDPIs. 
Specifically, a channel $P_{Y|X}$ is said to satisfy an SDPI with contraction coefficient  $\eta$ if
\begin{equation}
\eta I(U;X) \geq I(U;Y) \label{eq:SDPIdef}
\end{equation}
for every $P_{UX}$ such that $U\markov X\markov Y$ is a Markov chain. The minimal $\eta$ for which~(\ref{eq:SDPIdef})
holds is denoted by $\etakl(P_{Y|X})$, and is given by
\begin{equation}
\etakl(P_{Y|X}) = \sup\frac{I(U;Y)}{I(U;X)} \label{eq:sup_KL}
\end{equation}
where the supremum is taken over all $P_{UX}$ such that $I(U;X)>0$.
By their definitions,
\begin{equation}
\etakl(P_{Y|X}) = \etaln(\IchannX,P_{Y|X}) \label{eq:eta_Ichann}
\end{equation}
where $\IchannX$ stands for the identity channel with input alphabet ${\cal X}$.
In view of Lemma~\ref{lemma:LNequiv} and~(\ref{eq:eta_Ichann}) we have
\begin{corollary}
\label{cor:SDPI_equiv}
A channel $P_{Y|X}$ satisfies SDPI with constant $\eta$ if and only if
$\eta H(X) - I(X;Y)$ is concave in $P_X$.
\end{corollary}
The SDPI constant $\etakl$ is a property of a single channel,
measuring the dissipation of information over $P_{Y|X}$~\cite{PolyanskiyWu:17p}.
By contrast, $\etaln$ is a property
of a pair $\left\{ P_{Y_1|X}, P_{Y_2|X}\right\}$, measuring the dominance
of one channel over the other.
To cast $\etakl$ in the notation of $\etaln$~(\ref{eq:sup_LN}),
assume that the pair $\left\{ P_{Y_1|X}, P_{Y_2|X}\right\}$ is stochastically or physically degraded,
i.e. there exists a channel $P_{Y_2|Y_1}$ such that $P_{Y_2|X}=P_{Y_2|Y_1}\circ P_{Y_1|X}$.
Then we can write
\begin{equation}
\etakl(P_{Y_2|Y_1}) = \sup_{P_{UY_1}} \frac{I(U;Y_2)}{I(U;Y_1)}=  
\sup_{P_{UX}}\sup_{P_{Y_1|X}} \frac{I(U;Y_2)}{I(U;Y_1)}\label{eq:etakl_BC}
\end{equation} 
where the supremum is over all the conditional distributions $P_{Y_1|U}$ such that $I(U;Y_1)>0$.
Comparing~(\ref{eq:etakl_BC}) to~(\ref{eq:sup_LN}), we see the following differences
between the  two definitions:
\begin{enumerate}
\item Part of  the channel over which~(\ref{eq:etakl_BC}) is optimized, is held fixed in~(\ref{eq:sup_LN})
\item In~(\ref{eq:sup_LN}) we dispense with  the Markov structure $X\markov Y_1\markov Y_2$.
\end{enumerate}
The subscript kl is used in~(\ref{eq:sup_KL}) since the minimal $\eta$ 
can be expressed as a solution of Kullback-Leibler 
divergence optimization problem~\cite{PolyanskiyWu:17p},~\cite{AnantharamGohariKamathNair:13p}:
\begin{equation}
\etakl(P_{Y|X}) = \sup_{Q} \etat(Q) \label{eq:etakl_div1}
\end{equation}
where
\begin{equation}
\etat(Q) \eqdef \sup_{P:\ 0<D(P\parallel Q)<\infty} \frac{D(P_{Y|X}\circ P\parallel P_{Y|X}\circ Q)}{D(P\parallel Q)},\label{eq:etakl_div2}
\end{equation}
$P,Q$ are distributions on the input alphabet ${\cal X}$, and for notational convenience we dropped the dependence
of $\etat(Q)$ on $P_{Y|X}$. In Lemma~\ref{lemma:eta_relations} we state
a few properties of $\etaln$, along with a characterization that parallels~(\ref{eq:etakl_div1}-\ref{eq:etakl_div2}).
\begin{lemma}
\label{lemma:eta_relations}
For any strongly LN pair $\left\{ P_{Y_1|X}, P_{Y_2|X}\right\}$, the following hold:
\begin{enumerate}
\item 
\begin{equation}
\etaln(P_{Y_1|X}, P_{Y_2|X}) \geq \frac{\etakl(P_{Y_2|X})}{\etakl(P_{Y_1|X})}.\label{eq:ln_lower1}
\end{equation}
\item
If the pair is further stochastically degraded, then  
\begin{equation}
\etaln(P_{Y_1|X}, P_{Y_2|X}) \leq \etakl(P_{Y_2|Y_1}). \label{eq:ln_upper1}
\end{equation}
\item
$\etaln$ can be expressed as
\begin{equation}
\etaln(P_{Y_1|X},P_{Y_2|X}) = \sup_Q \etate(Q)  \label{eq:etaln_div1}
\end{equation}
where
\begin{equation}
\etate(Q) \eqdef 
           \sup \frac{D(P_{Y_2|X}\circ P\parallel P_{Y_2|X}\circ Q)}{D(P_{Y_1|X}\circ P\parallel P_{Y_1|X}\circ Q)}, \label{eq:etaln_div2}
\end{equation}
$P,Q$ are distributions on $\alphabetX$, and the sup in~(\ref{eq:etaln_div2}) is taken over all $P$ such that
\begin{equation}
0<D(P_{Y_1|X}\circ P\parallel P_{Y_1|X}\circ Q) <\infty .  \label{eq:etaln_div3}
\end{equation}
\end{enumerate}
\end{lemma}
The proof of Lemma~\ref{lemma:eta_relations}
is given in Section~\ref{subsec:proof_eta_relations}. The definition of $\etate(Q)$ in~(\ref{eq:etaln_div2}) 
is an extension of~(\ref{eq:etakl_div2}), to include a channel in the denominator, hence the superscript e.
\begin{remarks}[Input constraints and the Gaussian channel]
\label{remark:input_Constraints}
The definitions of $\etaln$ and $\etamc$
 and the bound~(\ref{eq:ln_lower1}) can be extended to include input constraints. 
In~\cite{CalmonPolyanskiyWu:18p,PolyanskiyWu:16p} 
it is shown that the Gaussian channel with input moment constraints do not satisfy
linear SDPI, i.e. $\etakl=1$. 
Hence~(\ref{eq:ln_lower1}) implies that $\etaln=1$ for the Gaussian BC.
As for SPDIs, non-linear degradedness relations are needed in order
to derive results for the Gaussian BC and PRC.
This is treated briefly in Section~\ref{sec:nonlinear}.
\end{remarks}

In the following examples we evaluate $\etaln$ for pairs of common channels,
and compare the results to the bounds in Lemma~\ref{lemma:eta_relations}. Proofs of the examples
are given in Appendix~\ref{appendix:examps_etas}.

\begin{example}[Binary erasure channels]
\label{examp:bec_bec}
Let $P_{Y_i|X}$ be a binary erasure channel with parameter $p_i$ (BEC($p_i$)), $i=1,2$, $0\leq p_1< p_2\leq 1$.
Then the pair is stochastically degraded, and
\begin{equation}
\etamc=\etaln =\frac{\olsi{p}_2}{\olsi{p}_1} = \frac{\etakl(\mbox{BEC}(p_2))}{\etakl(\mbox{BEC}(p_1))}
 \label{eq:etas_BEC}
\end{equation}
where $\olsi{p} = 1-p$. Thus, for a pair of erasure channels, $\etamc$ and $\etaln$ coincide,
and the lower bound in Lemma~\ref{lemma:eta_relations} is tight.
\end{example}

\begin{example}[Binary symmetric channels]
\label{examp:bsc_bsc}
In this example $P_{Y_i|X}$, $i=1,2$ are binary symmetric channels (BSCs) with parameters $p_i$, $0\leq p_1\leq p_2\leq 1/2$.
The pair of BSCs is stochastically degraded, i.e., $P_{Y_2|X} = P_{Y_2|Y_1}\circ P_{Y_1|X}$,
where $P_{Y_2|Y_1} = \mbox{BSC}((p_2-p_1)/(1-2p_1))$. We have
\begin{equation}
\etaln = \frac{(1-2p_2)^2}{(1-2p_1)^2} = \frac{\etakl(\mbox{BSC}(p_2))}{\etakl(\mbox{BSC}(p_1))}=\etakl(P_{Y_2|Y_1})\label{eq:etaln_BSC}
\end{equation}
and
 \begin{equation}
 \etamc = \max_{0\leq \theta\leq 1/2}\frac{H_b(\theta*p_2) - H_b(p_2)}{H_b(\theta*p_1) - H_b(p_1)} 
 = \frac{1-H_b(p_2)}{1-H_b(p_1)} , \label{eq:etamc_BSC} 
 \end{equation}
 where $\theta*p$ is the cyclic convolution $\theta*p=\olsi{\theta}p +\theta\olsi{p}$.
We conclude that  for a pair of BSCs, the lower and upper bounds of Lemma~\ref{lemma:eta_relations} are tight,
and $\etamc<\etaln$ for $0<p_1<p_2<1/2$. 
\end{example}

\begin{example}[Z channel and BSC]
\label{examp:z_bsc}
Consider now the pair $\left\{\mbox{Z}(\epsilon),\mbox{BSC}(p)\right\}$ 
where for the Z channel $P_{Y_1|X}(1|0) = \epsilon$, $P_{Y_1|X}(0|1) = 0$.
Due to the asymmetry of the Z channel, an exact expression
of $\etakl$ in  terms of $\epsilon,\theta$ is quite involved. 
It is solved numerically. Figure~\ref{fig:Z_BSC} depicts $\etaln$ and the ratio
$\etakl(\mbox{BSC}(p)/\etakl(\mbox{Z}(\epsilon)$ for $\epsilon=0.3$
and $0\leq p\leq 1/2$. The lower bound~(\ref{eq:ln_lower1}) is not tight 
in this example, and for $p<0.19$ the channel is not degraded.

\begin{figure}
    \centering
    \includegraphics[width=\linewidth]{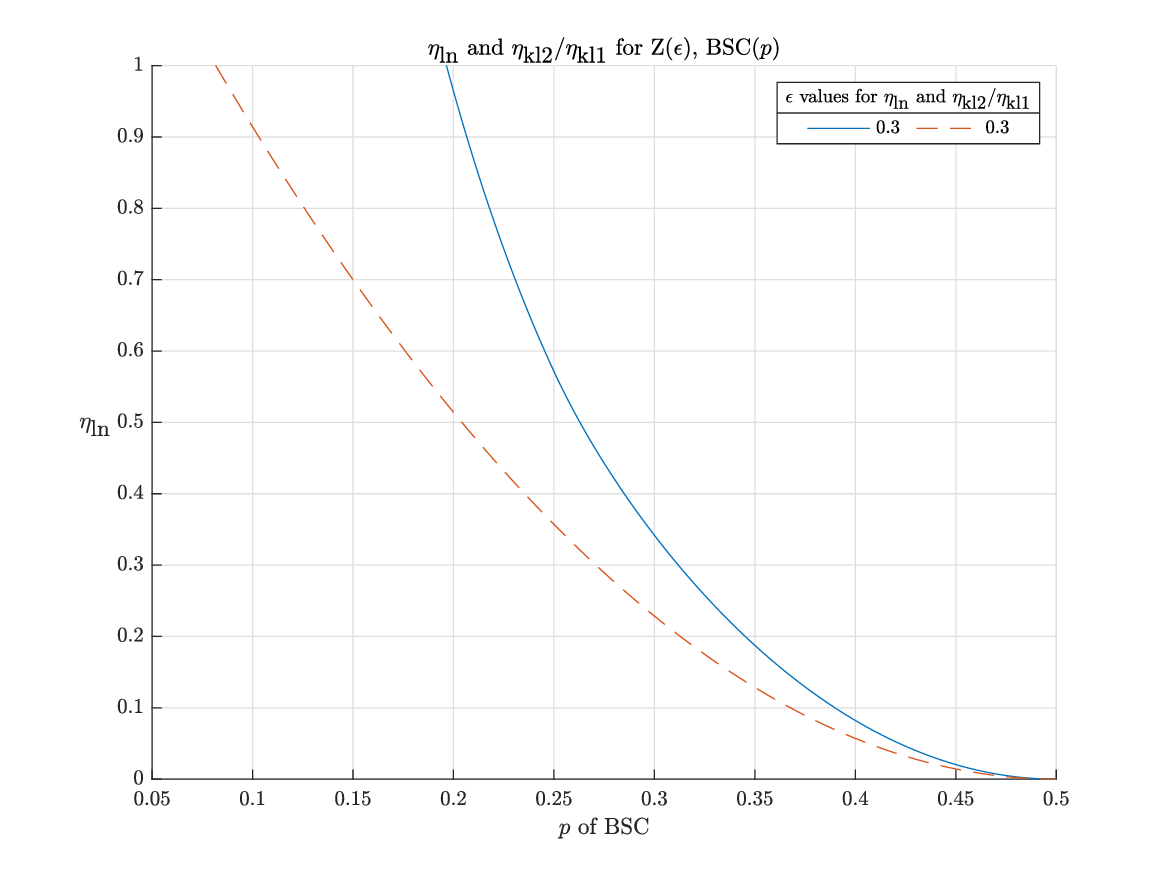}
    \caption{Curves of $\etaln$ (solid) and $\etakl(\mbox{BSC}(p))/\etakl(\mbox{Z}(0.3))$ (dashed). For $p<0.19$
    the channel is not degraded}
    \label{fig:Z_BSC}
\end{figure}

\end{example}
\begin{remarks}[Computation of $\etaln$, $\etakl$ and Inclusion Relations]
\label{remarks:inclusion}
\begin{enumerate}
\item
Lemma~\ref{lemma:LNequiv} provides a simple way to calculate $\etaln$: 
compute the Hessian of the functional $J(\eta,P_X) = \eta I(X;Y_1) - I(X;Y_2)$
w.r.t. $P_X$, and check for the minimal value of $\eta$ such that the Hessian is negative
semi-definite. 
These steps can be performed numerically, especially when the input alphabet is small.
In particular, for binary input channels, the Hessian is the second derivative
of $J(\eta,P_X) $ w.r.t. the parameter of a Bernoulli($\theta$) input distribution,
and computation of $\etaln$ for symmetric channels can be done analytically.
The same comment holds for the calculation of SDPI coefficients $\etakl$, according to
Corollary~\ref{cor:SDPI_equiv}.
\item
By the classical definitions of degraded, LN and MC relations, 
any degraded channel is also LN and MC. However, a degraded channel 
is not necessarily $\etalnset$ nor $\etamcset$ for any $\eta<1$. 
The simplest such example is a physically degraded BC with 
$P_{Y_2|Y_1}$ an identity channel.
\item 
For any $\eta'\geq\eta$, we have $\etalnset\subseteq\etalnsetp{\eta'}\subseteq\etamcsetp{\eta'}$.
A reverse inclusion does not hold: given a pair $\eta' > \eta$, in general $\etalnsetp{\eta'}\not\subseteq\etamcset$.
The simplest example is a pair of channels for which $\etaln=\etamc$, e.g.
a pair of BECs as in Example~\ref{examp:bec_bec}.
\end{enumerate}
\end{remarks}

\section{Cooperative BC}
\label{sec:main_BC}
In this section we present our result on the BC with conference link.
Let $P_{Y_1,Y_2|X}$ be a broadcast channel with input alphabet $\alphabetX$ and  output alphabets 
$\alphabetY_1$, $\alphabetY_2$. Observe that due to the cooperation, the capacity region depends, in general,
on the joint distribution of the outputs. 
Let $n$ be the transmission length, and $\alphabetN_k = [1:\nu_k]$  the message set for user $k$, 
of size $\nu_k$, $k=1,2$. A cooperation link from user 1 to user 2 is present, with capacity $C_{12}$.
Denote by $\alphabetN_c = [1:2^{nC_{12}}]$ the cooperation message set. 
\begin{definition}
\label{def:cooperative_BC}
An $(n,\nu_1,\nu_2,2^{nC_{12}},\epsilon)$ code for the cooperative broadcast channel
is an encoder
\begin{subequations}
\label{subeq:code_coopBC}
\begin{equation}
f:\alphabetN_1\times \alphabetN_2\rightarrow \alphabetX^n, \label{subeq:def_coopBC_enc}
\end{equation}
a cooperation encoder
\begin{equation}
f_c: \alphabetY_1^n\rightarrow \alphabetN_c\label{subeq:def_coopBC_enc2}
\end{equation}
and a pair of decoders:
\begin{IEEEeqnarray}{rCl}
& & \phi_1 :\alphabetY_1^n \rightarrow \alphabetN_1\label{subeq:def_coopBC_dec1}\\
& & \phi_2 :\alphabetY_2^n\times \alphabetN_c \rightarrow \alphabetN_2\label{subeq:def_coopBC_dec2}
\end{IEEEeqnarray}
\end{subequations}
with probability of error bounded by $\epsilon$:
\begin{equation}
\frac{1}{\nu_1\nu_2}\sum_{(j_1,j_2)\in\alphabetN_1\times\alphabetN_2} P_{Y_1^n,Y_2^n|X^n}\left(D^c_{j_1,j_2}|f(j_1,j_2)\right)\leq\epsilon
\label{eq:def_coopBC_Pe}
\end{equation}
where $D_{j_1,j_2}$ is the decoding set of  the message pair $(j_1,j_2)$
\begin{IEEEeqnarray}{rCl}
\IEEEeqnarraymulticol{3}{l}{
D_{j_1,j_2} = \left\{(y_1^n,y_2^n): \right. }\nonumber\\* \qquad \quad
& & \left. \left\{\phi_1(y_1^n)=j_1\right\}\cap \left\{ \phi_2(y_2^n,f_c(y_1^n)) = j_2\right\}\right\}
 \label{eq:def_coopBC_decset}
\end{IEEEeqnarray}
and $D^c_{j_1,j_2}$ its complement.
The rates  $R_1,R_2$ of the code are the pair
\begin{equation}
R_k=\frac{\log \nu_k}{n}, \ \ k=1,2. \label{eq:def_coopBC_rates}
\end{equation}
A rate pair $(R_1,R_2)$ is said to be achievable with cooperation capacity $C_{12}$ if for every $\delta>0$, $\epsilon>0$ 
and sufficiently large $n$, there exists an  $(n,2^{n(R_1-\delta)},2^{n(R_2-\delta)},2^{nC_{12}},\epsilon)$ code for  the channel.
The capacity region with cooperation link of capacity $C_{12}$ is the set of all achievable pairs $(R_1,R_2)$,
and is denoted by ${\set C}(C_{12})$.
\end{definition}

Let $\setRi(C_{12})$ be the set of all pairs $(R_1,R_2)$ satisfying
\begin{subequations}
\label{subeq:BC_i}
\begin{IEEEeqnarray}{rCl}
R_1 &\leq& I(X;Y_1|U) \label{subeq:BC_R1_i}\\
R_2 &\leq& I(U;Y_2) + C_{12} \label{subeq:BC_R2_i}\\
R_1+R_2 &\leq& I(X;Y_1) \label{subeq:BC_sum_i}
\end{IEEEeqnarray}
for some joint distribution $P_{U,X} P_{Y_1,Y_2|X}$.
\end{subequations}
The region $\setRi(C_{12})$ is achievable for any cooperative BC, i.e., 
\begin{equation}
\setRi(C_{12}) \subseteq {\set C}(C_{12}) \label{eq:Ri_in_C}
\end{equation}
and equals the capacity region for physically degraded channels
(see~\cite{LiangVeeravalli:07p},\cite{DaboraServetto:06p} and note that physical degradedness
is not used there in the direct part). The proof of~(\ref{eq:Ri_in_C}) is outlined in Section~\ref{subsec:proof_BC_main},
for completeness. 

Denote by $C_k$ the capacity of $P_{Y_k|X}$, i.e. $C_k=\max_{P_X} I(X;Y_k)$, $k=1,2$.
In our main result, stated next, 
we focus on channels with $C_2>0$.
\begin{theorem}
\label{theo:BC_main}
Let $P_{Y_1,Y_2|X}$ be a cooperative BC with $C_2>0$, $P_{Y_1|X}\geqln{\eta} P_{Y_2|X}$ and
\begin{subequations}
\label{subeq:C12_R2_constraints}
\begin{equation}
C_{12}\leq \frac{1-\etaln}{\etaln} C_2.\label{subeq:C12_constraint}
\end{equation}
Then $\setRi(C_{12})$ is tight for
\begin{equation}
R_2\geq \frac{C_{12}}{1-\etaln}. \label{subeq:R2_region}
\end{equation}
Furthermore, under~(\ref{subeq:C12_R2_constraints}), the inequalities~(\ref{subeq:BC_R1_i}) and~(\ref{subeq:BC_R2_i}) 
dominate~(\ref{subeq:BC_sum_i}) for any $U$, and the sum rate bound can be dropped.
\end{subequations}
\end{theorem}
The proof of Theorem~\ref{theo:BC_main} is given in Section~\ref{subsec:proof_BC_main}.
We make the following observations:
\begin{enumerate}
\item
The strong LN property and the characterization given in Theorem~\ref{theo:BC_main},
depend on the channel $P_{Y_1,Y_2|X}$ only via its conditional marginals $P_{Y_1|X},\ P_{Y_2|X}$.
\item 
The fact that D\&F is optimal means that under~(\ref{subeq:C12_R2_constraints})
the order of the users is kept.
\item
The r.h.s of~(\ref{subeq:C12_constraint}) (resp.~(\ref{subeq:R2_region})) is a monotone decreasing
(resp. increasing) function of $\etaln$. Thus, an upper bound on $\etaln$ suffices
to apply Theorem~\ref{theo:BC_main}, and there is no need to compute $\etaln$ exactly.
 In case that only upper bounds on $\etaln$ are available,
the penalty would be reduced ranges of $C_{12}$ and $R_2$ for which 
we know that D\&F is optimal.
\end{enumerate}

\begin{figure*}[!tbp]
\subfloat[\label{subfig:BEC_BSC_deg}] {\includegraphics[width=.49\linewidth]{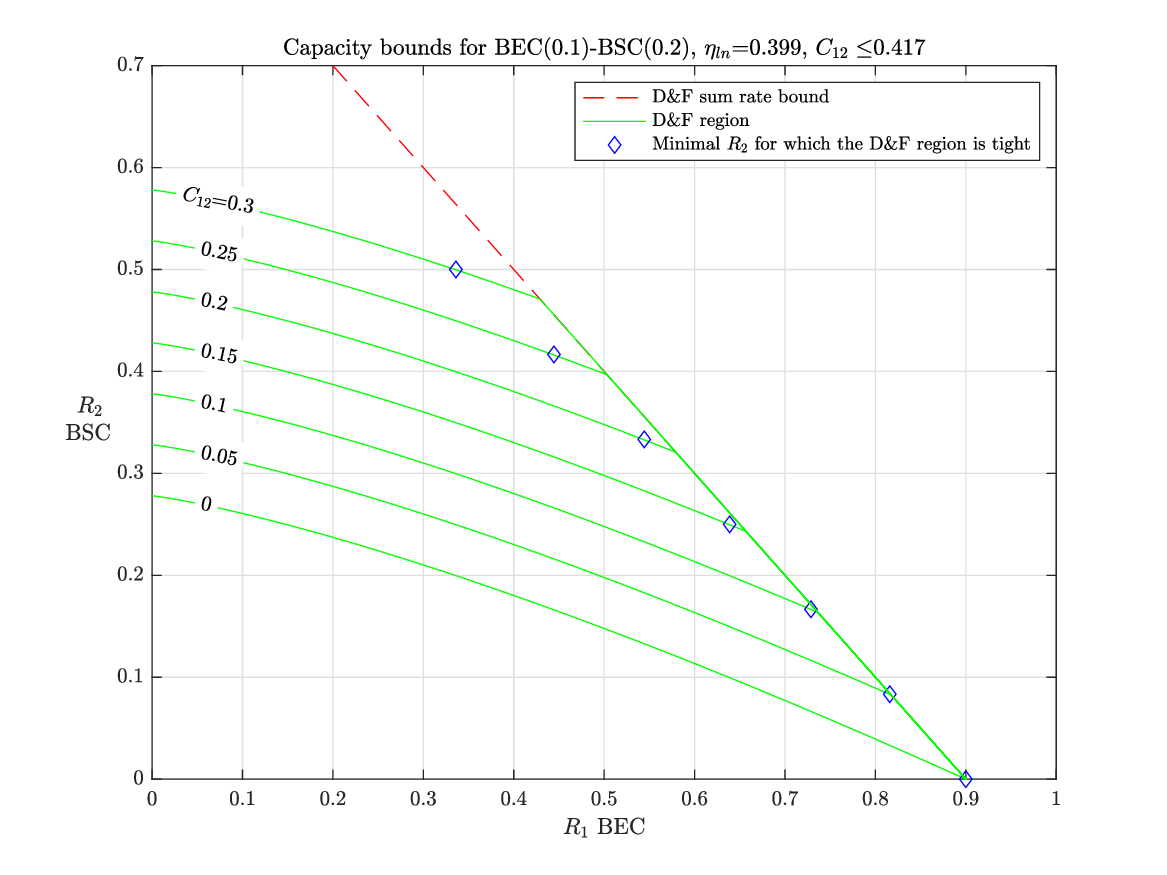}%
}\hfill
\subfloat[\label{subfig:BEC_BSC_nondeg}] {\includegraphics[width=.49\linewidth]{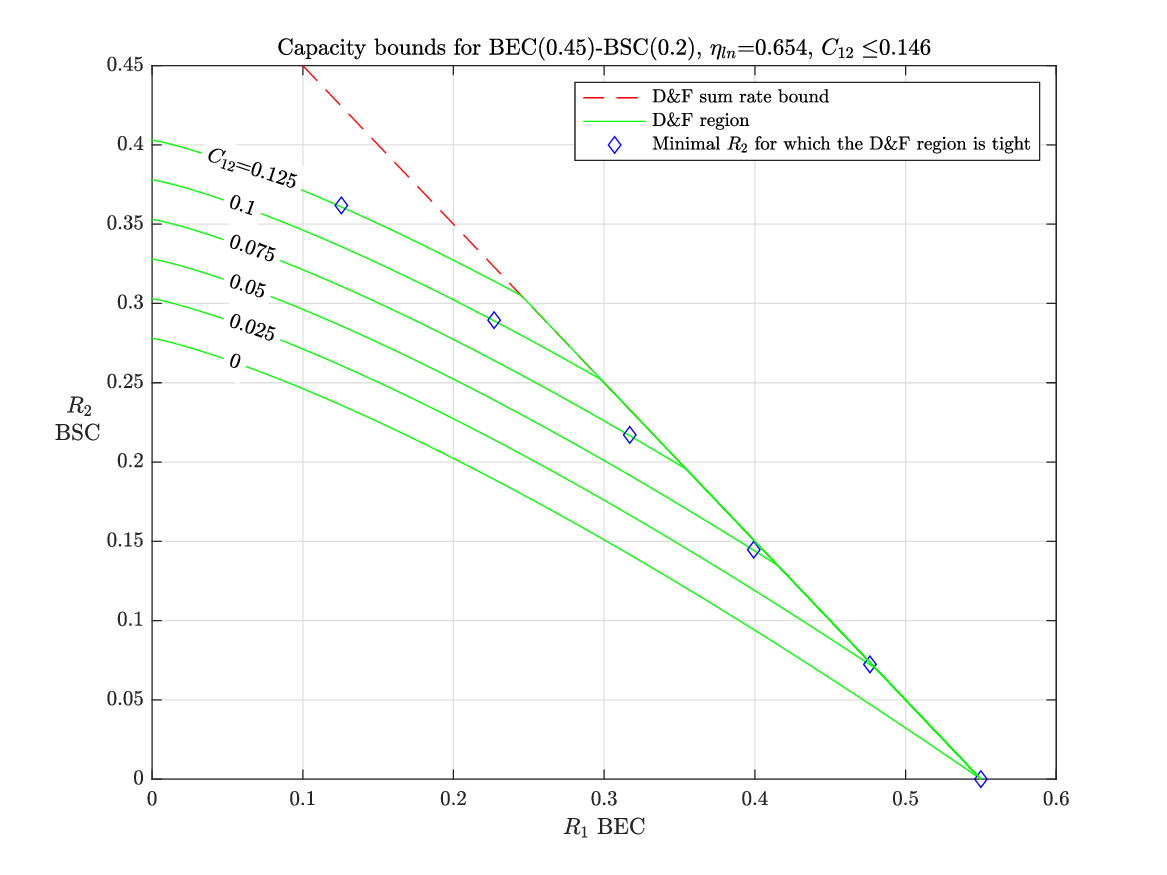}%
}
\caption{D\&F regions for the BEC($\epsilon$)-BSC($p$) BC, for few values of conference link capacities $C_{12}$.
The blue diamonds mark the minimal $R_2$ for which the D\&F region is tight, for the corresponding conference rate.
In Fig.~\ref{subfig:BEC_BSC_deg} $(\epsilon,p)=(0.1,0.2)$, the channel is stochastically degraded, 
thus it has a physically degraded version for which the D\&F region is tight for all values of $C_{12}$ and $R_2$.
In Fig.~\ref{subfig:BEC_BSC_nondeg} $(\epsilon,p)=(0.45,0.2)$ so the channel is less noisy but not 
stochastically degraded.}
\label{fig:BEC_BSC}
\end{figure*}

\begin{example}[BEC-BSC Broadcast Channel]
\label{examp:bec_bsc_cap}
In this example we compute the bounds developed in Theorem~\ref{theo:BC_main}
for the broadcast channel composed of a BEC($\epsilon$) (user 1) and BSC($p$) (user 2).
It is shown in~\cite{Nair:10p} that the channel is less noisy for $0\leq\epsilon\leq 4p(1-p)$,
and stochastically degraded for $0\leq\epsilon\leq 2p$. We show in Appendix~\ref{appendix:example_bec_bsc_cap}
that
\begin{equation}
\etaln = \frac{(1-2p)^2}{1-\epsilon}\quad \mbox{for} \ 0\leq\epsilon\leq 4p(1-p).\label{eq:eta_eamp_BC}
\end{equation}
The D\&F region ${\set R}_i(C_{12})$ (\ref{subeq:BC_i}) is the set of all pairs $(R_1,R_2)$ satisfying
\begin{subequations}
\label{subeq:examp_bec_bsc_cap}
\begin{IEEEeqnarray}{rCl}
R_1 &\leq& \bar{\epsilon}H_b(\alpha)\label{subeq:examp_bec_bsc_cap_R1}\\
R_2 &\leq &1-H_b(\alpha*p) + C_{12}\label{subeq:examp_bec_bsc_cap_R2}\\
R_1+R_2 &\leq& \bar{\epsilon}\label{subeq:examp_bec_bsc_cap_sumrate}
\end{IEEEeqnarray}
for some $0\leq\alpha\leq 1/2$. It is tight for $C_{12}$ and $R_2$ satisfying
\begin{IEEEeqnarray}{rCl}
C_{12} &\leq& \frac{4p(1-p)-\epsilon}{(1-2p)^2} (1-H_b(p)) \label{subeq:examp_bec_bsc_cap_C12_bound}\\
R_2 &\geq& \frac{1-\epsilon}{4p(1-p) -\epsilon} C_{12}. \label{subeq:examp_bec_bsc_cap_R2_bound}
\end{IEEEeqnarray}
\end{subequations}
Fig.~\ref{fig:BEC_BSC} depicts the region~(\ref{subeq:examp_bec_bsc_cap}) for a few values of $C_{12}$.
The blue diamonds mark the minimal value of $R_2$ for which the D\&F region coincides with the capacity region,
for the corresponding link capacity $C_{12}$. The proof of Example~\ref{examp:bec_bsc_cap}
is given in Appendix~\ref{appendix:example_bec_bsc_cap}.
\end{example}

\section{Primitive relay channels}
\label{sec:PRC}
The primitive relay channel (PRC) can be viewed as a cooperative BC, 
with link capacity $C_{12}$ from user 1 to user 2,
where the rate for user 1 is 0. Ie, user 1 is not required to decode any messages. 
Hence it is a special case of Definition~\ref{def:cooperative_BC}.
Denote its capacity by $\Crelay(C_{12})$. Using  Theorem~\ref{theo:BC_main}
with $R_1=0$, and ignoring the lower bound on $R_2$ (since we maximize it),
we conclude that for any PRC with $P_{Y_1|X}\geqln{\eta} P_{Y_2|X}$
and conference link satisfying~(\ref{subeq:C12_constraint}),
\begin{equation}
\label{eq:PRC_cap_ln}
\Crelay(C_{12}) = C_2+C_{12}.
\end{equation}
But we can do slightly better.
We can extend the range of conference capacities $C_{12}$ and class of channels,
by using the strong MC relations, instead of strong LN.
This is stated in the next theorem.
\begin{theorem}
\label{theo:PRC_main}
Let $P_{Y_1,Y_2|X}$ be a PRC with $P_{Y_1|X}\geqmc{\eta} P_{Y_2|X}$,
and conference link of capacity
\begin{equation}
\label{eq:C12_mc}
C_{12}\leq \frac{1-\etamc}{\etamc} C_2.
\end{equation}
Then
\begin{equation}
\label{eq:PRC_cap_mc}
\Crelay(C_{12}) = C_2+C_{12}.
\end{equation}
\end{theorem}
\emph{Proof:}\\
The D\&F lower bound on the capacity of the PRC is:
\begin{equation}
\Crelay \geq \max_{P_X} \min\left\{ I(X;Y_1), I(X;Y_2) +C_{12} \right\}\label{eq:PR_df_lb}
\end{equation}
(this lower bound follows from classical results, 
e.g~\cite{Kim:07c},~\cite{ElGamalKim:11b} or~\cite{MondelliHasaniUrbanke:19p}.
It can be deduced also from~(\ref{subeq:BC_i}), by ignoring $R_1$).
Then
\begin{IEEEeqnarray}{rCl}
I(X;Y_1) &\geq& \frac{1}{\etamc} I(X;Y_2)\nonumber\\
         & =&  I(X;Y_2) + \frac{1-\etamc}{\etamc} I(X;Y_2),\label{eq:PR_df_lb1}
\end{IEEEeqnarray}
hence
\begin{equation}
I(X^*;Y_1) \geq C_2+\frac{1-\etamc}{\etamc} C_2 \geq C_2 + C_{12} \label{eq:PR_df_lb2}
\end{equation}
where $X^*$ is the capacity achieving distribution of $P_{Y_2|X}$.
Therefore, $I(X;Y_2) +C_{12}$ is the dominating term in the r.h.s. of~(\ref{eq:PR_df_lb}).
It is tight by the cut-set bound.
\QEDB

\section{Diamond channels}
\label{sec:diamond}
In the general diamond channel (DC) model, a source is connected to the destination
via a $K$-users BC, followed by a $K$-users multiple access channel (MAC). The BC outputs are forwarded
to the MAC inputs via relays - units that are not required  to decode any messages,
and act only as intermediate nodes between the source and the destination.
The relays can either be non cooperating, or cooperating with given conference capacities.
Two main sub-models exist in the literature.
The MAC diamond channel (MDC) is a diamond channel whose broadcast 
 stage consists of orthogonal links with given finite capacities.
Similarly, the broadcast diamond channel (BDC) is a diamond channel 
whose MAC stage consists of orthogonal links
with given finite capacities. This channel can be viewed also as $K$ primitive relay channels
with common input and destination. A broadcast diamond channel with two non cooperating relays
is depicted in Fig.~\ref{fig:BDC2}.

In~\cite{SanderovichShamaiSteinbergKramer:08p} the two-relay BDC was studied, 
under the assumption of oblivious relays - i.e., relays that are not informed about the codebook
used by the main encoder,
thus refer to the coding scheme at the BC  input as a random code.  
Various models of MDC, with and without side information, were studied in~\cite{
LiSimeoneYener:13p,ZhaoDingKhisti:15p,BidokhtiKramer:16p,BidokhtiKramerShamai:17p,WangWiggerZaidi:18p,
DiksteinBidokhtiShamai:22c}.

\begin{figure}
\centering 
 {\scalefont{0.75}
  \begin{tikzpicture}[auto,node distance = 0.8cm and 0.8cm,scale=1,>={Kite[scale=0.6]},
    box_e/.style={rectangle,draw=black,inner sep=3pt,minimum height=0.5cm,minimum width=0.8cm,align=center},     
       box_c/.style={rectangle,draw=black,inner sep=2pt,minimum height=0.75cm,minimum width=1.1cm,align=center}, 
       box_d/.style={rectangle,draw=black,inner sep=3pt,minimum height=0.5cm,minimum width=0.8cm,align=center}, 
        box_relay/.style={rectangle,draw=black,inner sep=2pt,minimum height=0.5cm,minimum width=0.8cm,align=center}, 
      helper/.style={ellipse,draw=black,text=black,fill=cyan!20,minimum height=0.4cm,minimum width=0.8cm,align=center}, 
      relay/.style={ellipse,draw=black,text=black,fill=cyan!20,minimum height=0.5cm,minimum width=1cm,align=center}, 
       smallbox/.style={rectangle,draw=black,inner sep=2pt,minimum height=0.4cm,minimum width=0.4cm,align=center}, 
       smalldot/.style={circle,draw=black,fill=black,thick,inner sep=0pt,minimum size=0.5mm}                      
       ]

      \node (m) {};
      \node [box_e] (encoder) [right=  of m] {Encoder};
      \node [box_c,right =  of encoder] (channel) {Channel\\$P_{Y_1Y_2Y_3|X}$};
      \node [box_d,right =  2cm of channel]  (decoder) {Decoder};
      \node (relayline) at ($(channel.east)!0.6!(decoder.west)$){};
      \node [relay, above=of relayline](relay1){Relay 1};
       \node [relay, below=of relayline](relay2){Relay 2};
      \node [right=0.75cm of decoder] (hat_m){};
      \coordinate (y1_out) at ($(channel.north east)!0.5!(channel.east)$);
      \coordinate (y2_out) at ($(channel.south east)!0.5!(channel.east)$){};

      \draw [->] (encoder) to node {$X$} (channel);
      \draw [->] (channel) to node[near start] {$Y_3$} (decoder);
      \draw [->] (m) to node[near start] {$m$} (encoder);
      \draw [->] (decoder) to node[near end] {$\hat{m}$} (hat_m);
      \draw [->] (y1_out) to node {$Y_1$} (relay1);
      \draw [->] (y2_out) to node[swap] {$Y_2$} (relay2);
      \draw [->] (relay1) to node {$C_{13}$} (decoder.north);
      \draw [->] (relay2) to node[swap] {$C_{23}$} (decoder.south);

\end{tikzpicture}
}
\caption{The broadcast diamond channel, composed of two primitive relay channels with common input and destination.}
\label{fig:BDC2}
\end{figure}

In this section we focus on the BDC with two separated (i.e., non cooperating) relays, as depicted in Fig.~\ref{fig:BDC2}.
Partial result is given also for cooperating relays, in Part 2 of Theorem~\ref{theo:BDC_main} below.
A formal definition of capacity is omitted, for space considerations.
We first derive a lower bound on the capacity, based on binning of the messages 
and D\&F at both relays.
We then show that under suitable strong-MC conditions, the lower bound is tight,
and achieves the cut-set bound. Define
\begin{IEEEeqnarray}{rCl}
R_i &=& \max_{P_X}\min\left\{I(X;Y_1), I(X;Y_2) \right. ,\nonumber\\
       & &\mbox{} \qquad \left. I(X;Y_3) + C_{13} + C_{23}\right\}.
\label{eq:Ri_BDC}
\end{IEEEeqnarray}
Denote by $C_k$ the capacity of $P_{Y_k|X}$, i.e. $C_k = \max_{P_X} I(X;Y_k)$, $k=2,3$.
Our main result on the BDC is stated next.
\begin{theorem}
\label{theo:BDC_main}
\begin{enumerate}
\item For any BDC with conference link capacities $C_{13}$ and $C_{23}$, $R_i$ is achievable.
\item Assume that $C_3>0$, $P_{Y_k|X}\geqmc{\eta^{k3}}P_{Y_3|X}$ for $k=1,2$ and
\begin{subequations}
\label{subeq:BDC_cap_cond1}
\begin{IEEEeqnarray}{rCl}
C_{13}+C_{23} &\leq & \frac{1-\etamct}{\etamct} C_3,\label{subeq:BDC_cap_cond1_2}
\end{IEEEeqnarray}
where
\begin{equation}
\etamct= \max\left\{\etamc^{13},\etamc^{23}\right\}.\label{subeq:BDC_cap_cond1_1}
\end{equation}
Then $R_i$ is tight, and the capacity is given by
\begin{equation}
C= C_3+C_{13} + C_{23}. \label{subeq:BDC_cap_1}
\end{equation}
\end{subequations}
Moreover, the capacity is given by~(\ref{subeq:BDC_cap_1})
also when the two relays communicate directly with each other, 
with arbitrary link capacities, $C^{(r)}_{12}$ and  $C^{(r)}_{21}$.
\end{enumerate}
\end{theorem}
The proof of Theorem~\ref{theo:BDC_main} is given in Section~\ref{subsec:proof_BDC}.
\begin{remarks}
\label{rem:diamond}
\begin{enumerate}
\item The broadcast channel $P_{Y_1,Y_2,Y_3|X}$ need not be physically nor stochastically degraded.
\item Similarly to Theorems~\ref{theo:BC_main} and~\ref{theo:PRC_main}, the conditions
under which the results of Theorem~\ref{theo:BDC_main} hold, and the capacity characterizations,
depend on the channel $P_{Y_1,Y_2,Y_3|X}$ only via its conditional marginals.
\end{enumerate}
\end{remarks}

\section{Nonlinear degradedness conditions}
\label{sec:nonlinear}
In this section we define strong LN functions,
in the spirit of data processing functions of~\cite{PolyanskiyWu:17p,CalmonPolyanskiyWu:18p}.
The purpose is to show that a version of Theorem~\ref{theo:BC_main}
holds in cases where $\etaln=1$. Moreover, when $\etaln<1$, the use of 
nonlinear strong LN relations can yield larger range of $C_{12}$ and $R_2$
for which D\&F is optimal. 

The nonlinear SDPI function developed in~\cite{CalmonPolyanskiyWu:18p}
is basically an extension of the contraction coeficient to functions.
It turns out that in our case,  it is more convenient to work
with expansion, or domination relation. It can be viewed as an inverse
of the nonlinear SDPI function. For completeness,
we give a short overview of both approaches.

We start with an LN pair,  in the sense of~\cite{KornerMarton:77p1},
i.e. $P_{Y_1|X}\geqln{} P_{Y_2|X}$. Let $C_k$ be the capacity of 
$P_{Y_k|X}$, $k=1,2$.
To make the discussion relevant also for the Gaussian channel, we include
an input constraint. Let ${\set P}$ be a convex set of distributions
on ${\set X}$.
Let ${\set F}(C_1,C_2)$, ${\set G}(C_1,C_2)$ be collections of functions:
\begin{IEEEeqnarray}{rCl}
{\set F}(C_1,C_2)& = &\left\{ F: [0,C_1]\rightarrow [0,C_2], \quad \mbox{increasing}, \right. \nonumber \\*
  &  &  \quad \left. F(0) =0,\quad  F(t)\leq t \right\} \label{eq:coll_F}\\
{\set G}(C_1,C_2)& = &\left\{ G: [0,C_2]\rightarrow [0,C_1], \quad \mbox{increasing}, \right. \nonumber \\*
  &  &  \quad \left. G(0) =0,\quad  G(t)\geq t \right\}. \label{eq:coll_G}
\end{IEEEeqnarray}
Denote by ${\set F}^{\cap}(C_1,C_2)$
the set of all concave functions in ${\set F}(C_1,C_2)$.
Note that all the members of ${\set F}^{\cap}(C_1,C_2)$
are subadditive.
Similarly, denote by ${\set G}^{\cup}(C_1,C_2)$ the set of all
 convex functions in 
${\set G}(C_1,C_2)$. All the members of ${\set G}^{\cup}(C_1,C_2)$ are superadditive
\begin{definition}
\label{def:LN_deg_dom}
A pair $P_{Y_1|X},P_{Y_2|X}$ is said to satisfy strong LN property with degradation function
$F\in{\set F}(C_1,C_2)$ if
\begin{equation}
I(U;Y_2)\leq F(I(U;Y_1))\quad \forall P_{UX},\ P_X\in{\set P}. \label{eq:Fdeg_def}
\end{equation}
It is said to satisfy strong LN property with domination function $G\in{\set G}(C_1,C_2)$ if
\begin{equation}
I(U;Y_1)\geq G(I(U;Y_2))\quad \forall P_{UX},\ P_X\in{\set P}. \label{eq:Gdom_def}
\end{equation}
\end{definition}
The minimal degradation function $F_1\in{\set F}(C_1,C_2)$ is given by
\begin{equation}
F_1(P_{Y_1|X},P_{Y_2|X},{\set P},t) \eqdef \sup_{\substack{P_{UX}:\\ P_X\in{\set P}}} \left\{ I(U;Y_2):   
  I(U;Y_1)\leq t\right\}. \label{eq:def_F1}
\end{equation}
Following the discussion leading to~(\ref{eq:eta_Ichann}),
the SDPI function $F_I$ defined in~\cite{CalmonPolyanskiyWu:18p}
can be expressed as
\begin{equation}
F_I(t) = F_1(\IchannX,P_{Y_2|X},{\set P},t)\label{eq:FI_F1}
\end{equation}
where $\IchannX$ is the identity channel with input alphabet ${\set X}$.
From this point on, when clear from the context, we drop  the dependence
of $F_1$ on the pair $\left\{ P_{Y_1|X}, P_{Y_2|X}\right\}$  and on ${\set P}$.
The upper concave envelope of $F_1$ is defined as
\begin{IEEEeqnarray}{rCl}
\label{eq:F1c}
F_1^c(t)  &= & \inf\left\{f(t): f(t)\geq F_1(t),\quad f(t)\ \mbox{concave} \right\}\nonumber\\
      &\in& {\set F}^{\cap}(C_1,C_2)
\end{IEEEeqnarray}
and is given by~(\cite{PolyanskiyWu:17p}):
\begin{equation}
F_1^c(t) = \sup_{\substack{P_{VUX}:\\ P_X\in{\set P}}} \left\{ I(U;Y_2|V):   
  I(U;Y_1|V)\leq t\right\}. \label{eq:F1c_2}
\end{equation}
We next present the domination function that is basically an inverse of $F_1$.
The maximal domination function $G_1\in{\set G}(C_1,C_2)$
is given by
\begin{equation}
G_1(P_{Y_1|X},P_{Y_2|X},{\set P},t) \eqdef \inf_{\substack{P_{UX}:\\ P_X\in{\set P}}} \left\{ I(U;Y_1):   
  I(U;Y_2)\geq t\right\} \label{eq:def_G1}
\end{equation}
The lower convex envelope of $G_1$ is defined as
\begin{IEEEeqnarray}{rCl}
\label{eq:G1c}
G_1^c(t)  &=& \sup\left\{g(t): g(t)\leq G_1(t),\quad g(t)\ \mbox{convex} \right\}\nonumber\\
                &\in& {\set G}^{\cap}(C_1,C_2)
\end{IEEEeqnarray}
and is given by
\begin{equation}
G_1^c(t) = \inf_{\substack{P_{VUX}:\\ P_X\in{\set P}}} \left\{ I(U;Y_1|V):   
  I(U;Y_2|V)\geq t\right\}. \label{eq:G1c_2}
\end{equation}
For any domination function $G(t)$,define
\begin{equation}
\label{eq:G0_def}
G_0(t) = G(t) - t.
\end{equation} 
Our main result on nonlinear domination is the following
\begin{theorem}
\label{theo:nonlinear_main}
Let $P_{Y_1,Y_2|X}$ be a cooperative BC with $C_2>0$, domination function $G$
and
\begin{subequations}
\label{subeq:C12_R2_bounds}
\begin{equation}
C_{12}\leq G_0(C_2). \label{subeq:nonlinear_C12_bound}
\end{equation}
Assume that $G_0$ is strictly increasing. Then ${\set R}_i(C_{12})$ is tight for
\begin{equation}
R_2\geq G_0^{-1}(C_{12}) + C_{12} \label{subeq:nonlinear_R2_bound}
\end{equation} 
\end{subequations}
Furthermore, under~(\ref{subeq:C12_R2_bounds}), the inequalities~(\ref{subeq:BC_R1_i}) and~(\ref{subeq:BC_R2_i}) 
dominate~(\ref{subeq:BC_sum_i}) for any $U$, and the sum rate bound can be dropped.
\end{theorem}
The proof of Theorem~\ref{theo:nonlinear_main}
is quite similar to the proof of Theorem~\ref{theo:BC_main}.
A sketch of the proof is given in Section~\ref{subsec:proof_nonlinear_relations}.

\section{Proofs}
\label{sec:proofs}
\subsection{Proof of Lemma~\ref{lemma:eta_relations}}
\label{subsec:proof_eta_relations}
\underline{Part 1.} By the definitions of $\etaln$, $\etakl$ we can write
\begin{IEEEeqnarray}{rCl}
\frac{I(U;Y_2)}{I(U;X)} &\leq& \frac{\etaln(P_{Y_1|X},P_{Y_2|X}) I(U;Y_1)}{I(U;X)}\nonumber\\
                                   &\leq& \etaln(P_{Y_1|X},P_{Y_2|X}) \etakl(P_{Y_1|X}) \label{eq:ln_lower1_pf}
\end{IEEEeqnarray}
and Part~1 follows by taking the sup in the l.h.s. of~(\ref{eq:ln_lower1_pf}).

\noindent
The proof of Part 2 follows from~(\ref{eq:etakl_BC}) and~(\ref{eq:sup_LN}).

\noindent
\underline{Part 3.} The result for $\etakl$, (\ref{eq:etakl_div1}-\ref{eq:etakl_div2}), 
was proved in~\cite{AnantharamGohariKamathNair:13p} for the discrete case, 
and extended in~\cite{PolyanskiyWu:17p} for continuous alphabets.
The proof of Part~3 uses similar ideas, and is given here for completeness.

Given two distributions $P$, $Q$ on ${\cal X}$ and a channel $P_{U|X}$, we use the shorthand
\begin{IEEEeqnarray*}{rCl}
P_U &=& P_{U|X}\circ P\\
Q_U &=& P_{U|X}\circ Q\\
P_{Y_1} &=& P_{Y_1|X}\circ P\\
Q_{Y_1} &=& P_{Y_1|X}\circ Q\\
P_{Y_1|U} &=& \frac{\left(P_{Y_1|X}P_{U|X}\right)\circ P}{P_{U|X}\circ P}\\
Q_{Y_1|U} &=& \frac{\left(P_{Y_1|X}P_{U|X}\right)\circ Q}{P_{U|X}\circ Q}
\end{IEEEeqnarray*}
and similarly for $P_{Y_2}$ and $Q_{Y_2}$. Fix $Q$ that is not a point mass, and define 
\begin{equation}
\etalnt(Q)  \eqdef  \sup_{P_{U|X}} \frac{I(U;Y_2)}{I(U;Y_1)} = \sup_{P_{U|X}}
\frac{D(Q_{Y_2|U}\parallel Q_{Y_2}|Q_U)}{D(Q_{Y_1|U}\parallel Q_{Y_1}|Q_U)}\label{eq:etalnt_def}
\end{equation}
where $X$ is distributed according to $Q$. Then
\begin{IEEEeqnarray}{rCl}
I(U;Y_2) &=& D(Q_{Y_2|U}\parallel Q_{Y_2}|Q_U)\nonumber\\
               & \leq & \etate(Q) D(Q_{Y_1|U}\parallel Q_{Y_1}|Q_U)\nonumber\\
               &=& \etate(Q) I(U;Y_1). \label{eq:lemma3_oned}
\end{IEEEeqnarray}
Therefore
\begin{equation}
\etalnt(Q) \leq \etate(Q), \label{eq:etalntQ_leq_etateQ}
\end{equation}
implying
\begin{equation}
\etaln\leq \sup_{Q} \etate (Q). \label{eq:etaln_leq_etate}
\end{equation}
We now show that an inequality holds also in the opposite direction.
Fix $Q$, and choose $P$ such that
$D(P\parallel Q)<\infty$ and $0<D(P_{Y_1}\parallel Q_{Y_1})$. Let $U\sim\mbox{Bernoulli}(\epsilon)$,
and define
\begin{IEEEeqnarray}{rCl}
\tilde{Q} &=& \frac{Q-\epsilon P}{\bar{\epsilon}}\label{eq:Qt_def}\\
Q_{X|U=u} &=& \begin{cases}
P & \text{if } u=1\\
\tilde{Q} & \text{if } u=0.
\label{eq:pxu_def}
\end{cases}
\end{IEEEeqnarray}
Note that with~(\ref{eq:Qt_def})-(\ref{eq:pxu_def}), $Q_X=Q$. Thus
\begin{IEEEeqnarray}{rCl}
 \frac{I(U;Y_2)}{I(U;Y_1)} &=& \frac{D(Q_{Y_2|U}\parallel Q_{Y_2}|Q_U)}{D(Q_{Y_1|U}\parallel Q_{Y_1}|Q_U)}\nonumber\\
  &=& \frac{\bar{\epsilon} D(\tilde{Q}_{Y_2}\parallel Q_{Y_2}) + \epsilon D(P_{Y_2}\parallel Q_{Y_2})}
          {\bar{\epsilon} D(\tilde{Q}_{Y_1}\parallel Q_{Y_1}) + \epsilon D(P_{Y_1}\parallel Q_{Y_1})} .\label{eq:I_D}
 \end{IEEEeqnarray}
Taking now the limit of small $\epsilon$, we have
\begin{equation}
\lim_{\epsilon\rightarrow 0}\frac{\bar{\epsilon}}{\epsilon} D(\tilde{Q}_{Y_1}\parallel Q_{Y_1})
\stackrel{(a)}{\leq} \lim_{\epsilon\rightarrow 0}\frac{\bar{\epsilon}}{\epsilon} D(\tilde{Q}\parallel Q) \stackrel{(b)}{=}0 \label{eq:lim_div1}
\end{equation}
where $(a)$ follows from data processing inequality for divergence,
 and $(b)$ is one of the steps in~\cite[Theorem 4]{PolyanskiyWu:17p}. It can be verified by L'H{\^o}pital's rule.
 In a similar manner
 \begin{equation}
 \lim_{\epsilon\rightarrow 0}\frac{\bar{\epsilon}}{\epsilon} D(\tilde{Q}_{Y_2}\parallel Q_{Y_2}) = 0. \label{eq:lim_div2}
 \end{equation}
 Therefore, taking the supremum over $P_{U|X}$ in the l.h.s. of~(\ref{eq:I_D}), we conclude that
 \begin{equation}
 \etalnt(Q) \geq \frac{D(P_{Y_2}\parallel Q_{Y_2})}{D(P_{Y_1}\parallel Q_{Y_1})} \label{eq:lim_div3}
 \end{equation}
for every pair $P,Q$. The claim now follows by maximizing over $P,Q$.
\QEDB

\subsection{Proof of Theorem~\ref{theo:BC_main}}
\label{subsec:proof_BC_main}

\noindent
\subsubsection{Direct Part}
\label{subsubsec:main_BC_direct}
The proof of the direct part  is based on superposition coding and D\&F. This technique is
standard. An outline of the proof is given here for completeness. We use the definitions and $\delta$-convention
of~\cite{CsiszarKorner:82b}.

\noindent
\underline{Code construction}:
Pick a joint distribution $P_{UX}$. 
\begin{itemize}
\item Generate $2^{nR_2}$ codewords of length $n$, $\bml{u}(m_2)$, $m_2\in\alphabetN_2$, iid
according to $P_U$.
\item For every $m_2\in\alphabetN_2$, generate $2^{nR_1}$ codewords of length $n$, $\bml{x}(m_1|m_2)$
$m_1\in\alphabetN_1$, according to $\prod_{i=1}^n P_{X|U}(\cdot|u_i(m_2))$.
\item Divide the set $\alphabetN_2$ into $2^{nC_{12}}$ bins, each containing $2^{n[R_2-C_{12}]_+}$.
Denote by $l(m_2)$ the bin number in which $m_2$ resides.
\end{itemize}

\noindent
\underline{Encoding}. To transmit the pair $(m_1,m_2)$, the encoder sends $\bml{x}(m_1|m_2)$ via the channel.

\noindent
\underline{Decoding}. We start with user 1. Upon receiving the channel output $\bml{y}_1$, the decoder
looks for the unique pair $(j_1,j_2)$ such that
\begin{equation}
\left(\bml{u}(j_2), \bml{x}(j_1|j_2), \bml{y}_1\right) \in \alphabetT_{UXY_1}. \label{eq:main_BC_dec1}
\end{equation}
If such a pair exists and is unique, the decoder declares $(\hat{m}_1,\hat{m}_2)=(j_1,j_2)$. Otherwise,
an error is declared. Note that the decoder performs joint decoding of $m_1$ and $m_2$. Using standard results,
for this step to succeed with high probability it suffices to require
\begin{subequations}
\label{subeq:main_BC_DEC1_bounds}
\begin{IEEEeqnarray}{rCl}
R_1 & < & I(X;Y_1|U) \label{subeq:main_BC_DEC1_R1_ach_bounds}\\
R_1+R_2 &\leq & I(X;Y_1). \label{subeq:main_BC_DEC1_R1plusR2_ach_bounds}
\end{IEEEeqnarray}
\end{subequations}
 After decoding, the decoder sends the bin number $l(\hat{m}_2)$ via the conference link, to user 2. 
 
Assume that the decoding process of user 1 succeeded. Upon receiving the bin number $l(m_2)$
from user 1 via the conference link, and his output $\bml{y}_2$, the decoder looks in bin number $l(m_2)$ 
for the unique index $j_2$ such that
\begin{equation}
\left(\bml{u}(j_2),\bml{y}_2\right)\in \alphabetT_{U,Y}. \label{eq:main_BC_dec2}
\end{equation}
If such an index exists and is unique, the decoder declares that $\hat{m_2} = j_2$. Otherwise, and error is declared.
For this step to succeed with high probability it suffices to require
\begin{equation}
R_2 - C_{12} < I(U;Y_2). \label{eq:main_BC_DEC2_bound}
\end{equation}
The bounds~(\ref{subeq:main_BC_DEC1_bounds}),~(\ref{eq:main_BC_DEC2_bound}) coincide with~(\ref{subeq:BC_i}),
thus proving the achievability of $\setRi$.

\subsubsection{Converse Part}
\label{subsubsec:main_BC_converse} Assume we have a sequence
of $(n,2^{nR_1},2^{nR_2},2^{nC_{12}},\tilde{\epsilon}_n)_{n\geq 1}$
 codes for the cooperative BC with $\limn\tilde{\epsilon}_n=0$. 
Let $M_k$ stand for the random messages of user $k$,
$k=1,2$, and $Z^n$ for the random cooperation message sent from user 1 to user 2.
By the code definition, $H(Z^n)\leq nC_{12}$. Starting with Fano's inequality
\begin{subequations}
\label{subeq:BC_conv_Fano}
\begin{IEEEeqnarray}{rCl}
\IEEEeqnarraymulticol{3}{l}{
n(R_2-\epsilon_n) \leq I(M_2;Y_2^nZ^n)}\nonumber\\* \quad
 &=& I(M_2;Y_2^n) + I(M_2;Z^n|Y_2^n)\nonumber\\
            &\leq& I(M_2;Y_2^n) + H(Z^n)\nonumber\\
            & \leq& I(M_2;Y_2^n) + nC_{12}\label{subeq:BC_conv_FanoR2}
\end{IEEEeqnarray}
where $\limn\epsilon=0$. Similarly, we bound the sum rate
\begin{IEEEeqnarray}{rCl}
\IEEEeqnarraymulticol{3}{l}{
n(R_1+R_2-\epsilon_n) \leq I(M_1;Y_1^n|M_2) + I(M_2;Y^nZ^n)}\nonumber\\* \qquad
                               &\leq& I(M_1;Y_1^n|M_2) + I(M_2;Y^n) + nC_{12}. \label{subeq:BC_conv_FanoR1R2}
\end{IEEEeqnarray}
\end{subequations}
Following now the steps in the converse proof for the LN and MC broadcast channels
(e.g.,~\cite{KornerMarton:77p1},~\cite{ElGamal:79p}, see also~\cite[Sec. 5.6.1]{ElGamalKim:11b}),
we arrive to
\begin{subequations}
\label{subeq:U_i_bounds}
\begin{IEEEeqnarray}{rCl}
R_2- \epsilon_n &\leq & \one \sum_{i=1}^n I(U_i;Y_{2,i}) +C_{12} \label{subeq:U_i_bounds_R2}\\
R_1+R_2 &\leq& \one\sum_{i=1}^n \left[ I(X_i;Y_{1,i}|U_i) + I(U_i;Y_{2,i}) \right]\nonumber\\
& &\mbox{} +C_{12} \label{subeq:U_i_bounds_R1R2}
\end{IEEEeqnarray}
\end{subequations}
where
\begin{equation}
U_i = M_2Y_{1}^{i-1}Y_{2,i+1}^n. \label{eq:U_I_def.}
\end{equation}
Applying the classical time sharing argument, and taking the limit $n\rightarrow\infty$, we arrive at:
\begin{subequations}
\label{subeq:final_U_bounds}
\begin{IEEEeqnarray}{rCl}
R_2 &\leq& I(U;Y_2) +C_{12}\label{subeq:final_U_bounds_R2}\\
R_1+R_2 &\leq& I(X;Y_1|U) +I(U;Y_2) +C_{12}. \label{subeq:final_U_bounds_R1R2}
\end{IEEEeqnarray}
\end{subequations}
Next, we show that subject to a lower bound on the rate of user 2,~(\ref{subeq:final_U_bounds_R1R2})
implies a sum rate bound of the form of~(\ref{subeq:BC_sum_i}). 
Thus, let us focus on rates $R_2$ satisfying the conditions~(\ref{subeq:C12_R2_constraints}) of the theorem, 
which we repeat here for convenience:
\begin{subequations}
\label{subeq:C12_R2_constraints_proof}
\begin{equation}
R_2\geq \frac{C_{12}}{1-\etaln} = \frac{\etaln}{1-\etaln}C_{12} + C_{12} \label{subeq:R2_region_proof}
\end{equation}
and
\begin{equation}
C_{12}\leq \frac{1-\etaln}{\etaln} C_2.\label{subeq:C12_constraint_proof}
\end{equation}
\end{subequations}
Using~(\ref{subeq:R2_region_proof}) in the l.h.s of~(\ref{subeq:final_U_bounds_R2}) gives
\begin{equation}
\frac{\etaln}{1-\etaln}C_{12} + C_{12} \leq I(U;Y_2) + C_{12}, \label{eq:R2_region_proof2}
\end{equation}
hence 
\begin{equation}
I(U;Y_2)\geq \frac{\etaln}{1-\etaln} C_{12}. \label{eq:IUY2_bound}
\end{equation}
Note that~(\ref{subeq:C12_constraint_proof}) ensures that the r.h.s. of~(\ref{eq:IUY2_bound})
does not exceed $C_2$, the maximal value of $I(U;Y_2)$.
By the strong LN property, we can write
\begin{IEEEeqnarray}{rCl}
I(U;Y_1) &\geq& \frac{1}{\etaln}I(U;Y_2) = I(U;Y_2) + \frac{1-\etaln}{\etaln} I(U;Y_2)\nonumber\\
             & \stackrel{(a)}{\geq} & I(U;Y_2) + C_{12} \label{eq:IUY1IUY2_bound}
\end{IEEEeqnarray}
where in $(a)$ we used~(\ref{eq:IUY2_bound}). Substituting~(\ref{eq:IUY1IUY2_bound}) in~(\ref{subeq:final_U_bounds_R1R2})
results in the sum rate bound
\begin{IEEEeqnarray}{rCl}
R_1+R_2 &\leq& I(X;Y_1|U) +I(U;Y_1) = I(X;Y_1), \label{eq:sum_rate_bound_IUY1}
\end{IEEEeqnarray}
which is~(\ref{subeq:BC_sum_i}). 
Moreover, since~(\ref{subeq:BC_R1_i}) and~(\ref{subeq:BC_R2_i}) imply~(\ref{subeq:final_U_bounds_R1R2}),
we conclude that, subject to~(\ref{subeq:C12_R2_constraints_proof}),
the sum rate bound in $\setRi(C_{12})$ is dominated by the individual rate bounds. 
Therefore, it suffices to focus on the modified inner and outer bounds defined below,
that take into account the constraints~(\ref{subeq:C12_R2_constraints_proof}):

\noindent
\underline{The modified inner bound}~$\setRi'(C_{12})$ is the collection of all pairs $(R_1,R_2)$ satisfying
\begin{subequations}
\label{subeq:Ri_mod_bound} 
\begin{IEEEeqnarray}{rCl}
\frac{C_{12}}{1-\etaln} \leq R_2 &\leq& I(U;Y_2) + C_{12} \label{subeq:R2i_mod_bound}\\
R_1 &\leq& I(X;Y_1|U) \label{subeq:R1i_mod_bound}
\end{IEEEeqnarray}
\end{subequations}
for some joint distribution $P_{U,X} P_{Y_1,Y_2|X}$.

\noindent
\underline{The modified outer bound}~$\setRo'(C_{12})$ is the collection of all pairs $(R_1,R_2)$ satisfying
\begin{subequations}
\label{subeq:Ro_mod_bound} 
\begin{IEEEeqnarray}{rCl}
\frac{C_{12}}{1-\etaln} \leq R_2 &\leq& I(U;Y_2) + C_{12} \label{subeq:R2o_mod_bound}\\
R_1 + R_2&\leq& I(X;Y_1|U) + I(U;Y_2) + C_{12}\label{subeq:R1R2o_mod_bound}
\end{IEEEeqnarray}
\end{subequations}
for some joint distribution $P_{U,X} P_{Y_1,Y_2|X}$.

To conclude the proof of Theorem~\ref{theo:BC_main}, we have to show that $\setRi'(C_{12})=\setRo'(C_{12})$.
The direction $\setRi'(C_{12})\subseteq\setRo'(C_{12})$ is immediate. 
We show next that $\setRi'(C_{12})\supseteq\setRo'(C_{12})$. 
If $C_{12}$ equals its maximal value, then by~(\ref{subeq:C12_constraint_proof}) and (\ref{eq:R2_region_proof2})
we must have $I(U;Y_2)=C_2$, (\ref{subeq:Ri_mod_bound}), (\ref{subeq:Ro_mod_bound}) coincide,
and there is nothing to prove.
Thus assume that a strict inequality holds in~(\ref{subeq:C12_constraint_proof}).
In particular, it implies that there exist $\gamma>0$ and $P_{U,X}$ such that
\begin{equation}
\frac{C_{12}}{1-\etaln}  \leq I(U;Y_2) + C_{12}-\gamma. \label{eq:R2_mod_gamma}
\end{equation}
We have to show that for any $\gamma$ and $P_{U,X}$ satisfying~(\ref{eq:R2_mod_gamma}),
the point
\begin{subequations}
\label{subeq:line_gamma}
\begin{IEEEeqnarray}{rCl}
\frac{C_{12}}{1-\etaln}\leq R_2^{(o)} &=& I(U;Y_2) + C_{12} -\gamma\label{subeq:line_gamma_R2}\\
R_1^{(o)}+R_2^{(o)} &=& I(X;Y_1|U) + I(U;Y_2) + C_{12} \label{subeq:line_gamma_R1R2}
\end{IEEEeqnarray}
\end{subequations}
resides in the region
\begin{subequations}
\label{subeq:tilde_o_region}
\begin{IEEEeqnarray}{rCl}
\frac{C_{12}}{1-\etaln} \leq R_2 &\leq& I(\tilde{U};\tilde{Y}_2) + C_{12} \label{subeq:tilde_i_region_R2}\\
R_1 &\leq& I(\tilde{X};\tilde{Y}_1|\tilde{U}) \label{subeq:tilde_i_region_R1}
\end{IEEEeqnarray}
\end{subequations}
for some $P_{\tilde{U},\tilde{X}}$, where $\tilde{Y}_k$, $k=1,2$ are  the channel outputs with $\tilde{X}$ at the input.
Equivalently, we have to show that
\begin{subequations}
\label{eq:eq_points}
\begin{IEEEeqnarray}{rCl}
I(\tU;\tY_2) &\geq& I(U;Y_2)-\gamma\label{eq:eq_point_R2}\\
I(\tX;\tY_1|\tU) &\geq& I(X;Y_1|U) + \gamma.\label{eq:eq_point_R1}
\end{IEEEeqnarray}
\end{subequations}

Let $S$ be a random switch with $P_S(1) = \lambda$, $P_S(0) = \bar{\lambda}$. Define the random variables 
\begin{IEEEeqnarray}{rCl}
\dtU &=& \left\{ \,
           \begin{IEEEeqnarraybox}[] [c] {l?s}
           \IEEEstrut
           U & if $S=1$, \\
           \emptyset & if $S=0$,
           \IEEEstrut
           \end{IEEEeqnarraybox}
          \right.
     \label{eq:dtU_def}\\
\tU &=& (\dtU,S) \label{eq:tU_def}     
\end{IEEEeqnarray}
and the channel from $\tU$ to $\tX$:
\begin{IEEEeqnarray}{C}
P_{\tX|\tU} = \left\{ \, 
                    \begin{IEEEeqnarraybox}[] [c] {l?s}
                    \IEEEstrut
                    P_{X|U} & if $S=1$, \\
                    P_X       & if $S=0$.
                     \IEEEstrut
                     \end{IEEEeqnarraybox}
                     \right.
               \label{eq:Ptilde_def}
\end{IEEEeqnarray}
We have
\begin{subequations}
\label{subeq:Pt_I}
\begin{IEEEeqnarray}{rCl}
\label{subeq:Pt}
P_{\tX} &=& P_X\\
I(\tU;\tilde{Y}_2) &=& \lambda I(U;Y_2). \label{subeq:IY2t}
\end{IEEEeqnarray}
\end{subequations}
Choose now
\begin{equation}
\lambda = 1- \frac{\gamma}{I(U;Y_2)} \label{eq:gamma_choice}
\end{equation}
and note that by~(\ref{eq:R2_mod_gamma}), $\gamma\leq I(U;Y_2)$, hence~(\ref{eq:gamma_choice})
is a valid choice for $\lambda$. Using~(\ref{eq:gamma_choice}) in~(\ref{subeq:IY2t}) we get
\begin{equation}
I(\tU;\tilde{Y}_2) = I(U;Y_2) - \gamma, \label{eq:Itilde_value}
\end{equation}
proving~(\ref{eq:eq_point_R2}).

Turning to $R_1$, note that
\begin{IEEEeqnarray}{rCl}
I(\tX;\tY_1|\tU) &=& \lambda I(\tX;\tY_1|U,S=1) + \bar{\lambda}I(\tX;\tY_1|\emptyset,S=0)\nonumber\\
                        &=& \lambda I(X;Y_1|U) +\bar{\lambda} I(X;Y_1) \nonumber\\
                        &=& \lambda I(X;Y_1|U)\nonumber\\
                         & &\mbox{} + (1-\lambda)\left[I(X;Y_1|U) + I(U;Y_1)\right]\nonumber\\
                        &=& I(X;Y_1|U) +(1-\lambda) I(U;Y_1)\nonumber\\
                        &\stackrel{(a)}{=}& I(X;Y_1|U) + \gamma\frac{I(U;Y_1)}{I(U;Y_2)}\nonumber\\
                        &\geq& I(X;Y_1|U) + \gamma\label{eq:IXY1U}
\end{IEEEeqnarray}
proving~(\ref{eq:eq_point_R1}), where in $(a)$ we used~(\ref{eq:gamma_choice}). 
This completes the proof of the theorem.
\QEDB

\subsection{Proof of Theorem~\ref{theo:BDC_main}}
\label{subsec:proof_BDC}
\subsubsection{Achievability of $R_i$}
\label{subsubsec:BDC_achievability_proof}
Fix a rate $R$ to user 3.

\noindent
\underline{Code construction}:
Pick a distribution $P_X$.
\begin{itemize}

\item Generate $2^{nR}$ codewords of length $n$,  $\bml{x}(m)$,
$m\in \alphabetN\eqdef [1:2^{nR}]$ iid according to $P_X$.
\item
Divide the set $\alphabetN$ into $2^{nC_{23}}$ bins, each containing $2^{n[R-C_{23}]_+}$ messages.
Denote by  $l_2(m)$ the bin number in which $m$ resides,
and by ${\set B}_2(l_2(m))$ the bin content.
\item Further divide each bin ${\set B}_2(l_2)$, $l_2\in [1:2^{nC_{23}}]$, into $2^{nC_{13}}$ sub-bins,
each containing  $2^{n[R-C_{23}-C_{13}]_+}$ messages.
Denote by  $l_1(m)$ the number of the sub-bin in which $m$ resides,
and by ${\set B}_1(l_1(m),l_2(m))$ its content. Thus, for every $m\in\alphabetN$
\begin{equation}
m\in {\set B}_1(l_1(m),l_2(m)) \subseteq {\set B}_2(l_2(m)),\label{eq:BDC_proof_BinsDef}
\end{equation}
and 
\begin{IEEEeqnarray}{rCl}
\IEEEeqnarraymulticol{3}{l}{
\abs{ {\set B}_1(l_1,l_2) } \leq 2^{n[R - C_{13} - C_{23}]_+} }\nonumber\\* \qquad
& &\quad\forall\ \  l_1\in[1:2^{nC_{13}}],\quad l_2\in[1:2^{nC_{23}}]
\label{eq:BDC_proof_BinsDef_2}
\end{IEEEeqnarray}
\end{itemize}

\noindent
\underline{Encoding}: To transmit the message $m$, 
the encoder sends $\bml{x}(m)$ via the channel.

\noindent
\underline{Decoding}:
We start with relay node 2. The relay receives the signal $\bml{y}_2$.
He looks for an index $j_2$ such that
\begin{equation}
     (\bml{x}(j_2),\bml{y}_2)\in {\set T}_{XY_2}.\label{eq:BDC_proof_r2_1}
\end{equation}
If such index $j_2$ exists and is unique, relay 2 sets $\hat{m} = j_2$. Otherwise, an error is declared.
For this step to succeed with high probability, it suffices to require
 \begin{equation}
 R < I(X;Y_2). \label{eq:BDC_proof_r2_2}
 \end{equation}
 Relay 2 sends via the conference link $C_{23}$ the bin index in which $\hat{m}$ resides, $l_2(\hat{m})$.
 
Relay 1 decodes $m$. For this purpose, he looks for an index $j_1$
such that
\begin{equation}
(\bml{x}(j_1), \bml{y}_1) \in {\set T}_{XY_1}. \label{eq:BDC_proof_r1_1}
\end{equation}
If such pair exists, and is unique, the relay 1 sets $\dhat{m} = j_1$.
Otherwise, an error is declared.
For this step to succeed with high probability, it suffices to require
\begin{equation}
R < I(X;Y_1). \label{eq:BDC_proof_r1_2}
\end{equation}
Relay 1 then sends via the conference link $C_{13}$ the index of the sub-bin which $\dhat{m}$ resides,
$l_1(\dhat{m})$.

Assume from this point on that the decodings at the relays were successful.
Thus, the decoder gets the correct bin and sub-bin numbers, via the links $C_{13}$ and $C_{23}$.
Denote these indices by $l_1$ and $l_2$.
He then looks for an index $j$ in ${\set B}_1(l_1,l_2)$ such that
\begin{equation}
(\bml{x}(j), \bml{y}_3) \in {\set T}_{XY_3}.
\label{eq:BDC_proof_dest_1}
\end{equation}
If such index $j$ exists, and is unique, the decoder sets its decision
as $\tilde{m}=(j)$. Otherwise,
he declares an error. For this step to succeed with high probability, we require
 \begin{equation}
 R-C_{13}-C_{23} < I(X;Y_3). \label{eq:BDC_proof_r3}
 \end{equation}
Collecting~(\ref{eq:BDC_proof_r2_2}),~(\ref{eq:BDC_proof_r1_2}) and~(\ref{eq:BDC_proof_r3}),
we conclude that any  rate $R$ satisfying
\begin{IEEEeqnarray}{rCl}
R < \max_{P_{X}}\min && \left\{ I(X;Y_1), I(X;Y_2), \right. \nonumber\\
                                           &&           \left.  I(X;Y_3) + C_{13} + C_{23} \right\} .\label{eq:BDC_proof_final1}
\end{IEEEeqnarray}
is achievable.
This proves Part 1 of Theorem~\ref{theo:BDC_main}.

\subsubsection{Part 2 - capacity when $C_3>0$}
\label{subsubsec:BDC_cap_withC3}
We first compare the second and the third terms in~(\ref{eq:BDC_proof_final1}).
By~(\ref{subeq:BDC_cap_cond1_1}),~(\ref{subeq:BDC_cap_cond1_2}).
\begin{IEEEeqnarray}{rCl}
C_{13}+C_{23}&\leq& \min\left\{ \frac{ 1-\etamc^{23}}{\etamc^{23}}, \frac{ 1-\etamc^{13}}{\etamc^{13}}\right\}C_3
\nonumber\\
&=& \frac{ 1-\etamct}{\etamct} C_3. \label{eq:BCD_proof_linkConditions}
\end{IEEEeqnarray}
Then
\begin{equation}
I(X^*;Y_2)\geq \frac{1}{\etamc^{23}}C_3 =  C_3 + \frac{1-\etamc^{23}}{\etamc^{23}}C_3\geq C_3+C_{13}+C_{23},
\label{eq:BDC_proof_part2_1}
\end{equation}
where $X^*$ is the capacity achieving distribution of $P_{Y_3|X}$. Similarly,
\begin{equation}
I(X^*;Y_1)\geq \frac{1}{\etamc^{13}}C_3 = C_3 + \frac{1-\etamc^{13}}{\etamc^{13}}C_3\geq C_3+C_{13}+C_{23},
\label{eq:BDC_proof_part2_2}
\end{equation}
Therefore the third term in the minimum in~(\ref{eq:Ri_BDC}) dominate the first and the second terms.
The tightness follows from the cut set bound. 
Note that if the two relays can communicate with each other directly, the cut that achieves
$C_3+C_{13}+C_{23}$ is not affected by the links connecting them,
and the last statement of Part~2 follows.
\QEDB

\subsection{Proof of Theorem~\ref{theo:nonlinear_main}}
\label{subsec:proof_nonlinear_relations}
The proof follows the lines of the proof of Theorem~\ref{theo:BC_main}.
Likewise, we repeat the standard steps to conclude the inner bound ${\set R}_i(C_{12})$
of~(\ref{subeq:BC_i}) and the outer bound~(\ref{subeq:final_U_bounds}).
By the definition of $G_0(t)$, we have
\begin{equation}
\label{eq:proof_nonlinear_G0}
I(U;Y_1) \geq G_0(I(U;Y_2)) +I(U;Y_2).
\end{equation}
By~(\ref{subeq:final_U_bounds_R2}) and~(\ref{subeq:nonlinear_R2_bound})
\begin{equation}
I(U;Y_2) \geq G_0^{-1}(C_{12})\label{eq:proof_nonlinear_1}
\end{equation}
or
\begin{equation}
C_{12}\leq G_0(I(U;Y_2)). \label{eq:proof_nonlinear_2}
\end{equation}
Substituting~(\ref{eq:proof_nonlinear_2}) in~(\ref{subeq:final_U_bounds_R1R2}) we obtain
\begin{IEEEeqnarray}{rCl}
\label{eq:proof_nonlinear_3}
R_1+R_2 &\leq& I(X;Y_1|U) + I(U;Y_2) +C_{12}\nonumber\\
    &\leq& I(X;Y_1|U) + I(U;Y_2) +G_0(I(U;Y_2))\nonumber\\
    &\stackrel{(a)}{\leq}& I(X;Y_1|U) + I(U;Y_1) = I(X;Y_1) 
\end{IEEEeqnarray}
where in $(a)$ we used~(\ref{eq:proof_nonlinear_G0}).
Hence, the outer bound~(\ref{subeq:final_U_bounds}),
the conditions of Theorem~\ref{theo:nonlinear_main} (\ref{subeq:nonlinear_C12_bound}),
(\ref{subeq:nonlinear_R2_bound}) and~(\ref{eq:proof_nonlinear_G0})
imply a sum rate bound like~(\ref{subeq:BC_sum_i}).
Since~(\ref{subeq:BC_R1_i}) and~(\ref{subeq:BC_R2_i})
imply~(\ref{subeq:final_U_bounds_R2}) and~(\ref{subeq:final_U_bounds_R1R2}),
we conclude that the sum rate bound of ${\set R}_i(C_{12})$ is implied by the 
individual rate bounds.

We present now two modified bounds, parallel to~(\ref{subeq:Ri_mod_bound}) 
and~(\ref{subeq:Ro_mod_bound})

\noindent
\underline{The modified inner bound}~$\setRi'(C_{12})$ is the collection of all pairs $(R_1,R_2)$ satisfying
\begin{subequations}
\label{subeq:Ri_mod_bound_nonlinear} 
\begin{IEEEeqnarray}{rCl}
G_0^{-1}(C_{12}) + C_{12} \leq R_2 &\leq& I(U;Y_2) + C_{12} \label{subeq:R2i_mod_bound_nonlinear}\\
R_1 &\leq& I(X;Y_1|U) \label{subeq:R1i_mod_bound_nonlinear}
\end{IEEEeqnarray}
\end{subequations}
for some joint distribution $P_{U,X} P_{Y_1,Y_2|X}$.

\noindent
\underline{The modified outer bound}~$\setRo'(C_{12})$ is the collection of all pairs $(R_1,R_2)$ satisfying
\begin{subequations}
\label{subeq:Ro_mod_bound_nonlinear} 
\begin{IEEEeqnarray}{rCl}
G_0^{-1}(C_{12}) + C_{12}\leq R_2 &\leq& I(U;Y_2) + C_{12} \label{subeq:R2o_mod_bound_nonlinear}\\
R_1 + R_2&\leq& I(X;Y_1|U) + I(U;Y_2)\nonumber\\
& & \mbox{} + C_{12}\label{subeq:R1R2o_mod_bound_nonlinear}
\end{IEEEeqnarray}
\end{subequations}
for some joint distribution $P_{U,X} P_{Y_1,Y_2|X}$.
We have to show that the bounds are equivalent. The proof proceeds exactly along the lines
of the proof of equivalence of~(\ref{subeq:Ri_mod_bound}) and~(\ref{subeq:Ro_mod_bound}).
The details are omitted.
\QEDB

\appendix
\subsection{Derivation of Examples~\ref{examp:bec_bec},~\ref{examp:bsc_bsc},~\ref{examp:z_bsc}}
\label{appendix:examps_etas}
\noindent
\underline{Proof of Example~\ref{examp:bec_bec}}.
For BEC($p$) we have $I(X;Y) = \olsi{p} H_b(\theta)$
where $H_b$ is the binary entropy function, $X\sim\mbox{Bernoulli}(\theta)$ and $\bar{p}=1-p$.
Therefore
\begin{equation}  
\etamc =\sup_{P_X}\frac{I(X;Y_2)}{I(X;Y_1)} = \frac{\olsi{p}_2}{\olsi{p}_1}. \label{eq:etamc_BECs}
\end{equation}
By Corollary~\ref{cor:SDPI_equiv}, $\etakl$ is the maximal $\eta$ for which $\eta H_b(\theta) - \olsi{p}H_b(\theta)$
is concave in $\theta$. Since $H_b(\theta)$ is concave in $\theta$, we conclude that $\etakl=\olsi{p}$. 
Similarly, by Lemma~\ref{lemma:LNequiv},
$\etaln$ is the maximal $\eta$ for which $\eta\olsi{p}_1 H_b(\theta) - \olsi{p}_2 H_b(\theta)$ is concave in $\theta$.
Therefore 
\begin{equation}
\etaln = \frac{\olsi{p}_2}{\olsi{p}_1} = \frac{\etakl(\mbox{BEC}(p_2))}{\etakl(\mbox{BEC}(p_1))}=\etamc.\label{eq:etaln_BEC}
\end{equation}
\QEDB

\noindent
\underline{Proof of Example~\ref{examp:bsc_bsc}}.
For BSC,
\begin{equation}
I(X;Y_i) = H_b(\theta*p_i) - H_b(p_i) \label{eq:I_BSC}
\end{equation}
where $\theta*p$ is the cyclic convolution $\theta*p=\olsi{\theta}p +\theta\olsi{p}$.
From Corollary~\ref{cor:SDPI_equiv} we conclude that
\begin{equation}
\etakl(\mbox{BSC}(p)) = (1-2p)^2. \label{eq:etakl_BSC}
\end{equation}
By Lemma~\ref{lemma:LNequiv}, $\etaln$ is the minimal $\eta\leq 1$ such that
\begin{equation}
J(\eta,\theta) \eqdef \eta( H_b(\theta*p_2) - H_b(p_2) ) -(H_b(\theta*p_1) - H_b(p_1)) \label{eq:BSCln1}
\end{equation}
is concave in $\theta$.  To guarantee concavity of $J(\eta,\theta)$, 
we require that its second derivative w.r.t $\theta$ is negative for every $\theta$.
This results in the condition:
\begin{equation}
\eta \geq \frac{(1-2p_2)^2}{(1-2p_1)^2} A(\theta,p_1,p_2)
  \ \ \ \forall\ 0\leq\theta\leq 1/2 \label{eq:ln_BSC_cond}
\end{equation}
where
\begin{equation}
A(\theta,p_1,p_2) =  \frac{(\theta*p_1)(\overline{\theta*p_1})}{(\theta*p_2)(\overline{\theta*p_2})} .  
\label{eq:ln_BSC_cond_A}
\end{equation}
It is easy to verify that for $0\leq p_1\leq p_2\leq 0.5$ the maximal value of $A(\theta,p_1,p_2)$ is 1, 
attained at $\theta=1/2$. Therefore, using also~(\ref{eq:etakl_BSC})
\begin{equation}
\etaln = \frac{(1-2p_2)^2}{(1-2p_1)^2} = \frac{\etakl(\mbox{BSC}(p_2))}{\etakl(\mbox{BSC}(p_1))}.
\end{equation}
The last equality in~(\ref{eq:etaln_BSC}) 
 is obtained by substituting $p=(p_2-p_1)/(1-2p_1)$ in~(\ref{eq:etakl_BSC}).
  Finally, for $\etamc$, the maximum in~(\ref{eq:etamc_BSC}) is attained
 at $\theta=0.5$ (this is verified numerically).
 \QEDB

\noindent
\underline{Proof of Example~\ref{examp:z_bsc}}.
For the $Z$ channel
\begin{equation}
I(X;Y_1) = H_b(\olsi{\theta}\olsi{\epsilon}) - \olsi{\theta}H_b(\epsilon), \label{eq:I_Z}
\end{equation}
hence the function $J(\eta,P_X)$ is given by:
\begin{equation}
 J(\eta,\theta) = \eta\left[ H_b(\olsi{\theta}\olsi{\epsilon}) - \olsi{\theta}H_b(\epsilon)\right]
   -H_b(\theta*p) + H_b(\theta). \label{eq:J_Z_BSC}
\end{equation}
As in example~\ref{examp:bsc_bsc}, we take the second derivative of $J(\eta,\theta)$ w.r.t. $\theta$
and require it to be negative for every $\theta$. Therefore:
\begin{IEEEeqnarray}{rCl}
\ln(2)\frac{\partial^2 J(\eta,\theta)}{\partial \theta^2} &=& - \frac{\eta\olsi{\epsilon}^2}{\olsi{\theta}\olsi{\epsilon}(1-\olsi{\theta}\olsi{\epsilon})}
+\frac{(1-2p)^2}{(\theta*p)(\olsi{\theta*p})}\leq 0\nonumber\\
   & & \quad \forall\ 0\leq\theta\leq 1,
 \label{eq:J_2_dev}
 \end{IEEEeqnarray}
 and the condition on $\eta$ is
 \begin{equation}
 \eta\geq \frac{(1-2p)^2}{\olsi{\epsilon}}\frac{\olsi{\theta}(\theta\olsi{\epsilon}+\epsilon)}{(\theta*p)(\olsi{\theta*p})}\quad \forall\ 0\leq\theta\leq 1.
 \label{eq:eta_cond_Z_BSC}
 \end{equation}
 If, for some pair $(\epsilon,p)$, (\ref{eq:eta_cond_Z_BSC}) does not hold for  $\eta= 1$,
 then $\mbox{Z}(\epsilon)\notgeqln{\eta}\mbox{BSC}$ for any $\eta\leq 1$ and, in particular, $\mbox{Z}(\epsilon)$ is not less noisy
 than the $\mbox{BSC}(p)$. Since degradedness implies less noisy relations, 
 we also conclude that the BSC is not degraded  relative to the Z channel,
 with these parameters.
 Define
\begin{subequations}
\label{subeq:Z_BSC_K}
\begin{IEEEeqnarray}{rCl}
K(\theta,\epsilon,p) &\eqdef& \frac{\olsi{\theta}(\theta\olsi{\epsilon} +\epsilon)}{(\theta*p)(\overline{\theta*p})}
\label{eq:Z_BSC_K2}\\
K^* &\eqdef& \max_{0\leq \theta\leq 1} K(\theta,\epsilon,p).\label{subeq:Z_BSC_K1}
\end{IEEEeqnarray}
\end{subequations}
If
\begin{equation}
K^*\leq \frac{\olsi{\epsilon}}{(1-2p)^2},
\end{equation}
then~(\ref{eq:eta_cond_Z_BSC}) holds for some $\eta\leq 1$, $\mbox{Z}(\epsilon) \geqln{\eta}\mbox{BSC}(p)$, and
\begin{equation}
\etaln = \frac{(1-2p)^2}{\olsi{\epsilon}}K^* = \frac{\etakl(\mbox{BSC}(p))}{\etakl(\mbox{Z}(\epsilon))} K^*.
\label{eq:etaln_Z_BSC}
\end{equation}
Fig.~\ref{fig:Z_BSC} depicts $\etaln$ (computed numerically) and the ratio $\etakl(\mbox{BSC}(p))/\etakl(\mbox{Z}(\epsilon))$
for $\epsilon=0.3$ and $0\leq p\leq 1/2$. It  is seen that for $p\leq 0.19$ and $\epsilon=0.3$ the channel is not degraded.
Moreover, the lower bound of Lemma~\ref{lemma:eta_relations} is not tight.
\QEDB

\subsection{Proof of Example~\ref{examp:bec_bsc_cap}}
\label{appendix:example_bec_bsc_cap}
We start with the proof of~(\ref{eq:eta_eamp_BC}).
Let $X\sim P_X$ with $P_X(1)=\theta$, and define
\begin{IEEEeqnarray}{rCl}
J(\eta,\theta) &\eqdef& \eta I(X;Y_1) - I(X;Y_2)\nonumber\\
              &=& \eta\bar{\epsilon}H_b(\theta) - H_b(\theta * p) + H_b(p).\label{eq:examp_bec_bsc_proof_eta}
\end{IEEEeqnarray}
Then
\begin{equation}
\ln 2 \frac{\partial^2}{\partial\theta^2} J(\eta,\theta) = \frac{(1-2p)^2}{(\theta*p)\left(\olsi{\theta*p}\right)} - 
\frac{\eta\bar{\epsilon}}{\theta\bar{\theta}}
\label{eq:examp_bec_bsc_proof_der2}
\end{equation}
so $J(\eta,\theta)$ is concave in $\theta$ for
\begin{equation}
\eta\geq \frac{(1-2p)^2}{\bar{\epsilon}} A(\theta,0,p)\label{eq:examp_bec_bsc_proof_etaCond}
\end{equation}
where $A(\theta,p_1,p_2)$ is given in~(\ref{eq:ln_BSC_cond_A}) and its maximal value is~1,
proving~(\ref{eq:eta_eamp_BC}).

We proceed to show~(\ref{subeq:examp_bec_bsc_cap}). First, observe that
for $C_{12}=0$, the region~(\ref{subeq:BC_i}) coincides with~\cite[Eq.~(7)]{Nair:10p},
and  is evaluated in Claim~2 there. The distributions
that achieve the boundary of the capacity region have the marginal $P_X(1) = P_X(0) = 1/2$,
which also maximizes the sum rate bound. Therefore 
the additive term $C_{12}$ in~(\ref{subeq:BC_R2_i}) does not affect the maximization
of the information functions, and the capacity region is given 
by~(\ref{subeq:examp_bec_bsc_cap_R1},\ref{subeq:examp_bec_bsc_cap_R2},\ref{subeq:examp_bec_bsc_cap_sumrate}).
The bounds~(\ref{subeq:examp_bec_bsc_cap_C12_bound},\ref{subeq:examp_bec_bsc_cap_R2_bound})
are immediate, using~(\ref{eq:eta_eamp_BC}) in Theorem~\ref{theo:BC_main}.
This completes the proof of Example~\ref{examp:bec_bsc_cap}.
\QEDB

 \vspace{0.2cm}

\IEEEtriggeratref{7}

\bibliographystyle{hieeetr}
\bibliography{./references_Yossi_2025_09_16.bib}

\end{document}